\documentclass[10pt]{article}
\usepackage[utf8]{inputenc}
\usepackage[T1]{fontenc}
\usepackage[english]{babel}

\usepackage{geometry}
\usepackage[dvipdfmx]{graphicx}
\usepackage{xcolor}
\usepackage[capposition=top]{floatrow}

\usepackage{booktabs}
\usepackage{threeparttable}

\usepackage{mathtools,amssymb,mathrsfs,bm}

\allowdisplaybreaks
\DeclareMathOperator*{\argmax}{arg\,max}
\DeclareMathOperator*{\argmin}{arg\,min}

\DeclareMathOperator*{\esssup}{ess\,sup}
\DeclareMathOperator{\supp}{\mathrm{supp}}
\DeclareMathOperator{\mca}{\mathcal{A}}
\DeclareMathOperator{\mcg}{\mathcal{G}}
\DeclareMathOperator{\mcs}{\mathcal{S}}

\DeclareMathOperator{\mcx}{\mathcal{X}}
\DeclareMathOperator{\mcy}{\mathcal{Y}}
\DeclareMathOperator{\mcz}{\mathcal{Z}}

\DeclareMathOperator{\msb}{\mathscr{B}}
\DeclareMathOperator{\msp}{\mathscr{P}}
\DeclareMathOperator{\msu}{\mathscr{U}}

\DeclareMathOperator{\mbs}{\mathbb{S}}
\DeclareMathOperator{\mbt}{\mathbb{T}}
\DeclareMathOperator{\mbu}{\mathbb{U}}

\usepackage{enumitem}

\usepackage{xr}
\makeatletter%
\newcommand*{\addFileDependencyMAIN}[1]{
  \typeout{(#1)}%
  \@addtofilelist{#1}%
  \IfFileExists{#1}{}{\typeout{No file #1.}}%
}%
\makeatother%
\newcommand*{\myexternaldocumentMAIN}[1]{%
    \externaldocument[appendix-]{#1}%
    \addFileDependencyMAIN{#1.tex}%
    \addFileDependencyMAIN{#1.aux}%
}%
\myexternaldocumentMAIN{supp}%

\usepackage{xcolor}
\usepackage{hyperref}
\hypersetup{colorlinks=true,linkcolor=black,urlcolor=black,citecolor=black}

\usepackage{amsthm}

\theoremstyle{plain}
\newtheorem{theorem}{Theorem}[section]
\newtheorem{lemma}{Lemma}[section]
\newtheorem{corollary}{Corollary}[section]

\theoremstyle{definition}

\newtheorem{example}{Example}[section]
\newtheorem{assumption}{Assumption}[section]
\theoremstyle{remark}
\newtheorem{remark}{Remark}[section]

\usepackage[capitalize,nameinlink,noabbrev,nosort]{cleveref}
\crefname{theorem}{Theorem}{Theorems}
\Crefname{theorem}{Theorem}{Theorems}
\crefname{lemma}{Lemma}{Lemmas}
\Crefname{lemma}{Lemma}{Lemmas}
\crefname{corollary}{Corollary}{Corollaries}
\Crefname{corollary}{Corollary}{Corollaries}
\crefname{proposition}{Proposition}{Propositions}
\Crefname{proposition}{Proposition}{Propositions}
\crefname{assumption}{Assumption}{Assumptions}
\Crefname{assumption}{Assumption}{Assumptions}
\crefname{equation}{}{}
\Crefname{equation}{}{}
\creflabelformat{equation}{(#2#1#3)}

\setlist[enumerate]{label=(\roman*)}
\newlist{assuenum}{enumerate}{1} 
\setlist[assuenum]{label=(\roman*), ref=\theassumption.(\roman*)}
\crefalias{assuenumi}{assumption}
\newlist{corenum}{enumerate}{1}
\setlist[corenum]{label=(\roman*), ref=\thecorollary.(\roman*)}
\crefalias{corenumi}{corollary}
\newlist{lemenum}{enumerate}{1}
\setlist[lemenum]{label=(\roman*), ref=\thelemma.(\roman*)}

\usepackage[sort,longnamesfirst]{natbib}
\usepackage{layouts}

\usepackage{subfiles}

\newcommand{\OVERALL}{} 

\title{Distributionally Robust Policy Learning with Wasserstein Distance\thanks{The data used in this study are derived from data files made available to researchers by the MDRC. The author remains solely responsible for how the data have been used or interpreted. I am grateful to Xiaohong Chen, Riccardo D'Adamo, Takanori Ida, Toru Kitagawa, Shosei Sakaguchi, J{\"o}rg Stoye, and Takahide Yanagi for their helpful comments. I also thank the anonymous referees and the seminar participants at the 2021 Kansai Econometrics Meeting, 2022 IAAE, 2022 AMES, and 2022 SETA. Finally, this work was supported by a Grant-in-Aid for JSPS Fellows, Grant Number 19J20984.}}
\author{Daido Kido%
    \thanks{Graduate School of Economics, Kyoto University, Yoshida Honmachi, Sakyo, Kyoto, 606-8501, Japan. E-mail: \href{mailto:kido.daidou.45m@st.kyoto-u.ac.jp}{\texttt{kido.daidou.45m@st.kyoto-u.ac.jp}}.}}
\date{\empty}

\begin{document}

\begin{titlepage}
    \maketitle
    \begin{abstract}
    The effects of treatments are often heterogeneous, depending on the observable characteristics, and it is necessary to exploit such heterogeneity to devise individualized treatment rules (ITRs). Existing estimation methods of such ITRs assume that the available experimental or observational data are derived from the target population in which the estimated policy is implemented. However, this assumption often fails in practice because of limited useful data. In this case, policymakers must rely on the data generated in the source population, which differs from the target population. Unfortunately, existing estimation methods do not necessarily work as expected in the new setting, and strategies that can achieve a reasonable goal in such a situation are required. This study examines the application of distributionally robust optimization (DRO), which formalizes an ambiguity about the target population and adapts to the worst-case scenario in the set. It is shown that DRO with Wasserstein distance-based characterization of ambiguity provides simple intuitions and a simple estimation method. I then develop an estimator for the distributionally robust ITR and evaluate its theoretical performance. An empirical application shows that the proposed approach outperforms the naive approach in the target population. \\
    \vspace{0in}
    \bigskip
    \end{abstract}
\end{titlepage}

\newpage

\section{Introduction}\label{sec:introduction}
The effects of treatments are often heterogeneous based on observable covariates. In the presence of multiple treatments, an important decision for a policymaker is to choose a rule that specifies a treatment for each value of covariates to maximize their objective function. Such a rule is often called an \emph{individualized treatment rule (ITR)}, and its significance has been acknowledged in many areas including healthcare \citep{Bertsimas2017}, homeless services \citep{Kube2019}, and energy conservation \citep{Ida2021}.\par
The typical decision-making process of a policymaker is as follows. They are interested in a specific population, called the \emph{target population}. Population here refers to the joint distribution of potential outcomes and covariates. In determining an ITR, they can use experimental or observational data generated from another population, which is referred to as the \emph{source population}. Note that they can identify they source covariate distribution and source distribution of potential outcome associated with each treatment conditional on a possible value of covariates from this data. Based on this data, they decide upon an ITR, and then, the resulting ITR is applied to the target population. Finally, the average outcome attained by the ITR is realized as the policymaker's reward. The average outcome is often called \emph{welfare}; therefore, a policymaker's primal goal is to choose an optimal ITR that maximizes the target population's welfare.\par
Most existing studies have developed efficient estimation methods of optimal ITRs under the implicit assumption that the target and source populations are essentially identical. Under this assumption, the target population's welfare attained by arbitrary ITRs can be point-identified. Consequently, the optimal ITR for the target population also becomes identifiable. Based on this fact, previous studies have proposed an estimation method of optimal ITR utilizing the inverse-probability weighting \citep[e.g.,][]{Kitagawa2018,Swaminathan2015,Zhao2012} and augmented inverse-probability weighting \citep[e.g.,][]{Athey2021,Dudik2011,Zhou2022}.\par
However, in practice, whether this assumption is valid is controversial. For example, imagine that a state government wants to decide whether or not to allow each individual to participate in an employment support program to improve average earnings in the state. The available data on the program come from a randomized controlled trial conducted in another state. In the framework above, two available treatments exist: one is to force an individual to participate in the program, and the other is to exclude the individual from the program. Associated with each treatment, the potential outcome is earnings that would be realized when the individual is assigned to the treatment. In addition, observable individual attributes, such as age and previous earnings, compose covariates. The joint distribution of potential outcomes and covariates in the former state corresponds to the target population, and that in the latter state corresponds to the source population. The critical question is whether the former and latter states can be considered the same. The demographics of the two states may be different. In addition, the response to the program may also be different because of differences in other economic circumstances. In such cases, it would be unrealistic to assume that the two populations are the same. The scenarios in which the assumption can be violated are not limited to the above example. For instance, when the state government wants to exploit a randomized controlled trial conducted earlier in its state to determine the current optimal ITR for the state, it must consider the validity of the assumption judiciously.\par
\par
When the source and target populations are different, it is challenging to estimate the optimal ITR via the straightforward application of existing methods. However, in some cases, a slight modification in existing methods is sufficient to estimate the optimal ITR for the target population. Suppose that a difference exists between the source and target populations only in terms of covariate distribution and the density ratio of the target and source covariate distributions is known or identifiable with additional covariates data from the target population. In this case, the target population's welfare attained by an ITR remains identifiable by reweighting the outcome with the density ratio of source and target covariate distributions. \citet[][Remark~2.2]{Kitagawa2018} and \citet{Uehara2020} discuss estimation methods that are tailored to this situation. Instead, if the distributions of covariates or distributions of potential outcomes associated with each treatment conditional on a value of covariates differ between the source and target populations in an unknown and unidentifiable way, it is impossible to estimate the optimal ITR for the target population. In the first place, the target population's welfare attained by an ITR cannot be point-identified, which is a contrast to the case where the two populations are the same. This implies that the optimal ITR for the target population cannot be identified; the estimation goal itself is unclear.
\par
Motivated by this identification problem in such a situation, one strand of recent literature has proposed replacing the unidentifiable goal with an identifiable one. In particular, \citet{Mo2021}, \citet{Si2020,Si2021}, and \citet{Zhao2019} consider utilizing the concept of \emph{distributionally robust optimization (DRO)} to obtain a reasonable learning goal. DRO begins with constructing an \emph{ambiguity set}, which is a set of populations, based on the available knowledge and additional assumptions regarding the relationship between the source and target populations. Then, for each feasible ITR, it evaluates the \emph{distributionally robust welfare}, which is the worst-case value of welfare over the ambiguity set. Finally, it chooses the ITR that maximizes the distributionally robust welfare. Following \citet{Mo2021}, this study refers to such an ITR as a \emph{distributionally robust ITR (DR-ITR)}. By choosing the DR-ITR, it is guaranteed that the target population's welfare is at least tantamount to its distributionally robust welfare, as long as the target population belongs to the ambiguity set. From this perspective, the DR-ITR would be a reasonable goal.
\par
Although these attempts are the same, when they are applications of DRO, they vary based on underlying assumption. \citet{Zhao2019} and \citet{Mo2021} assume that the target population is absolutely continuous with respect to the source population and that a difference in the source and target populations exists only in the covariate distribution. Then, with the available data being the experimental or observational data generated from the source population, they construct the ambiguity set on the unknown target covariate distribution. Instead, \citet{Si2020,Si2021} include the case where the conditional distributions can also differ in the two populations, while assuming that the target population is absolutely continuous with respect to the source population. Using the experimental or observational data, which are derived from the source population, they construct the ambiguity set on the target population. As is clear from the current discussion, these attempts may not be appropriate when the assumed absolute continuity does not hold.
\par
This study also aims to define a reasonable ITR when the target population's welfare is not identifiable by exploiting the concept of DRO. Specifically, this study mainly deals with the situation where the policymaker has access to the experimental data from the source population and the covariate data from the target population. In other words, the policymaker has no knowledge about the target conditional distribution of potential outcomes. In contrast to the aforementioned attempts \citep{Mo2021,Zhao2019,Si2020,Si2021}, this study does not assume that the target conditional distribution of potential outcomes is absolutely continuous with respect to that of the source population, while the absolute continuity of the covariate distributions is maintained. To construct the ambiguity set without absolute continuity, I employ \emph{Wasserstein distance} or order 1. This is a metric over probability distributions that does not require the Radon-Nykodim derivative, and thus, the proposed method can cover broader scenarios.
\par
In addition to this advantage, the proposed Wasserstein-style ambiguity set is computationally attractive. In particular, with this ambiguity set, one can obtain the analytical solution of the distributionally robust welfare. Owing to this analytical simplicity, one can draw a lot of intuition. For example, it can be shown that obtaining an ITR by replacing the unknown target conditional distribution of potential outcomes with that of the source population is already distributionally robust under special cases in the Wasserstein sense. This provides a justification for the assumption that the source and target conditional distributions of potential outcomes are the same. In addition, by utilizing the technique developed in \citet{Mo2021} or \citet{Zhao2019}, one can easily extend the proposed method to a situation where no covariate data are available from the target population.
\par
For the estimation of the DR-ITR, the derived closed-from of the distributionally robust welfare implies a simple estimator. I assess its theoretical performance in terms of regret, which is in line with the literature on policy learning. Finally, I demonstrate the proposed DR-ITR using data from experimental evaluations of the changes in welfare-to-work programs in the early 1980s. The results show that the DR-ITR attains higher welfare in the target population than the ITR obtained by a naive approach.
\paragraph{Related Literature}
This study contributes to the literature on ITRs \citep[e.g.,][]{Kitagawa2018,Zhao2012,Swaminathan2015,Athey2021,Dudik2011,Zhou2022}. The methods developed in existing studies are intended for use in situations where the source and target populations are the same. The policy learnings when the two populations are different are not well-explored. This study designs a DR-ITR that works appropriately in such situations.
\par
The fundamental problem that motivates this study is the so-called \emph{external validity} \citep{Campbell1957}. The studies that explore external validity include \citep{Hotz2005,Cole2010,Stuart2011,Stuart2015,Stuart2018,Allcott2015,Andrews2019,Dehejia2021,Pritchett2013,Hartman2015,Gechter2019,Vivalt2020,Tipton2013,Tipton2014}. The primary interest of the literature is to obtain a point identification of the average treatment effect for the target population. To achieve this goal, the literature often assumes that the conditional distributions of potential outcomes are the same between the source and target populations. By contrast, this study allows a difference in the conditional distributions, and DRO attempts to obtain the lower bound of the target population's welfare.
\par
As mentioned earlier, the idea of this study stems from DRO with a Wasserstein-style ambiguity set. For a comprehensive review of DRO, see \citet{Rahimian2019}. A typical application of Wasserstein-DRO constructs an ambiguity set around the empirical distribution and lets its radius shrink to zero depending on the sample size to capture sampling uncertainty \citep[e.g.,][]{Esfahani2018,Blanchet2019,ShafieezadehAbadeh2015,ShafieezadehAbadeh2019,Gao2017a,Gao2020}. By contrast, the ambiguity set considered in this study is not centered around empirical distribution, but is centered around the source population, which differs from empirical distribution. In addition, its radius is set at a fixed positive value and does not converge to zero.
\par
\citet{Adjaho2022} conducted a similar study.\footnote{This study and \citet{Adjaho2022} are independent contributions made public almost simultaneously in the arXiv repository.} They also apply Wasserstein-DRO to a similar situation to develop a method that works when the source and target populations are different. One apparent difference is a concrete form of the Wasserstein-style ambiguity set. Specifically, \citet[][Section~3]{Adjaho2022} deals with a similar situation as that in this study, but the derived results are different due to the difference in the ambiguity set. Their result implies that the modifications discussed in \citet[][Remark~2.2]{Kitagawa2018} and \citet{Uehara2020} are already distributionally robust, while the result of this study implies that the distributional robustness of the modifications does not necessarily hold and this study's proposed approach outperforms the modifications in some aspects. In addition, \citet[][Section~4]{Adjaho2022} extends Wasserstein-DRO to the case where both the target covariate distribution and target conditional distribution of potential outcomes are unknown and unidentifiable, without imposing the assumption of the absolute continuity of the covariate distributions. Their result implies that the resulting distributionally robust welfare remains unidentifiable without further assumptions. By contrast, this study extends the proposed method to such a case while maintaining the assumption of the absolute continuity of the covariate distributions, and adopts the DRO technique developed in \citet{Mo2021,Zhao2019}.   
\paragraph{Structure of the paper}
In \cref{sec:model}, I formally introduce the underlying model considered in this study. Then, in \cref{sec:DRITR}, I discuss the application of Wasserstein-DRO to the model of \cref{sec:model}. \cref{sec:estimation} develops the estimation method of the proposed DR-ITR and provides the theoretical guarantee of the estimator. Finally, \cref{sec:empirical-application} demonstrates the proposed DR-ITR using experimental data. The additional information and proofs of the theoretical results are presented in the supplementary materials.
\paragraph{Notation}
For any metric space $(\mathcal{S},d_{\mathcal{S}})$, I use $\mathscr{B}(\mathcal{S})$ to denote Borel $\sigma$-algebra, which is the $\sigma$-algebra generated from the metric topology of $\mathcal{S}$. In addition, I write $\mathscr{P}(\mathcal{S})$ for the set of all probability measures on $(\mathcal{S},\mathscr{B}(\mathcal{S}))$. For a probability measure $\mathbb{P} \in \mathscr{P}(\mathcal{S})$, the support of the measure is denoted by $\supp(\mathbb{P})$. Given a measure space $(\mcs,\mathscr{B}(\mathcal{S}),\mathbb{P})$, another measurable space $(\mathcal{T},\mathscr{B}(\mathcal{T}))$, and a measurable map $f:(\mathcal{S},\mathscr{B}(\mathcal{S}))\mapsto(\mathcal{T},\mathscr{B}(\mathcal{T}))$, I denote the induced probability measure on $\mathscr{B}(\mathcal{T})$ by $f_{\#}\mathbb{P}$; that is, $f_{\#}\mathbb{P}(A)=\mathbb{P}(f^{-1}(A))$ for all $A\in\mathscr{B}(\mathcal{T})$. In addition, if $(\mathcal{T},\mathscr{B}(\mathcal{T})) = (\mathbb{R},\mathscr{B}(\mathbb{R}))$, the expectation of $f$ with respect to the measure $\mathbb{P}$ is denoted by $\mathbf{E}_{\mathbb{P}}[f]$.
\section{Model}\label{sec:model}
Here, I formally introduce the underlying model considered in this study (\cref{sec:problem-setup}). After introducing the model, I explain the naive approach that assumes that the two populations differ only in the covariate distributions (\cref{sec:naive-approach}). This approach often works as a bench mark. Before delving into details, a note of caution: \cref{sec:model,sec:DRITR} exclude the consideration of identification and estimation and focus on the problem of interest in the population level. In particular, I often assume that a policymaker has the knowledge of a certain population, which is a joint distribution of potential outcomes and covariates, although the joint distribution of potential outcomes is never identified even with the experimental data. The identification and estimation will be discussed in \cref{sec:estimation}.
\subsection{Problem Setup}\label{sec:problem-setup}
Let $\mca=\{1,\cdots,d\}$ be a finite set of possible treatments. Each individual in a population is characterized by a tuple $(Y_{1},\cdots,Y_{d},X)$, where $Y_{a}$ denotes the potential outcome that would be realized if one received treatment $a\in\mca$. I assume that for any $a\in\mca$, $Y_{a}$ takes a value in the outcome space $\mcy$, which is a closed subset of a Euclidean space. Then, a tuple $(Y_1,\cdots,Y_d)$ takes values in $\mcy^d$, which is assumed to be equipped with the $\ell_1$-distance; that is, for any pair $(y_1,\cdots,y_d),(y_1',\cdots,y_d') \in \mcy^d$, the distance is measured by $\sum_{a \in \mca} |y_a - y_a'|$. The last component $X$ denotes the individual's observable characteristics, which take value in a Polish space $\mcx$. Let $\mcz := \mcy^d \times \mcx$; thus, a tuple $(Y_1,\cdots,Y_d,X)$ takes values in $\mcz$. A particular population is characterized by a probability measure over $(\mcz,\msb(\mcz))$. For an arbitrary population $\mbu \in \msp(\mcz)$, let $\mbu_{(Y_1,\cdots,Y_d)\mid X=x} \in \msp(\mcy^d)$ denote the conditional distribution of potential outcomes, given by $X=x$, and let $\mbu_X \in \msp(\mcx)$ denote the marginal distribution of $X$.\footnote{In this study, the conditional distributions of potential outcomes are understood as a regular conditional probability measure. The existence of a regular conditional probability measure is guaranteed by \citet[Theorem~10.4.14]{Bogachev2007} because the covariate space is Polish. Hence, any conditional distribution of potential outcomes can be regarded as a proper probability measure in the outcome space, as compared to the case where one defines a conditional probability as a Radon-Nikodym derivative.} Additionally, I define the \emph{conditional mean response (CMR)} function as $m_{\mbu}(x,a):=\mathbf{E}_{\mbu_{(Y_{1},\cdots,Y_{d})|X=x}}[Y_{a}]$ for $a\in\mca$ and $x \in \supp(\mbu_X)$.\par
An ITR is a measurable mapping from $\mcx$ to $\mca$, and a set of candidate ITRs is denoted by $\mcg$. As emphasized in \citet{Kitagawa2018} and often assumed in subsequent studies, the set $\mcg$ may not necessarily contain all measurable mappings. For any ITR $g \in \mcg$, the welfare of ITR $g$ in population $\mbu \in \msp(\mcz)$ is given by
\begin{equation}
	V(g; \mbu) 
	:= \mathbf{E}_{\mbu}\left[\sum_{a \in \mca} Y_a \cdot 1\{g(X) = a\}\right]\\
	= \mathbf{E}_{\mbu_X}\left[\sum_{a \in \mca} m_{\mbu}(X,a) \cdot 1\{g(X) = a\}\right],
	\label{eq:welfare}
\end{equation}
where the last equality comes from the law of iterated expectations. Hence, the welfare of ITR $g$ can be calculated using the knowledge of the covariate distribution and CMR function of population $\mbu$. Clearly, the welfare of ITR $g$ varies with the population under consideration.\par
A policymaker focuses on the target population $\mbt\in\msp(\mcz)$, and aims to maximize the target population's welfare. Namely, given the target population $\mbt$, their goal would be to obtain the optimal ITR $g_{\mbt}^*$ for the target population as summarized in the following optimization problem:
\begin{equation}
    g_{\mbt}^* \in \argmax_{g \in \mcg} V(g; \mbt).
    \label{eq:standard-goal}
\end{equation}
It is obvious that the policymaker can achieve the goal if they have complete knowledge about $\mbt$. More precisely, knowledge about the target covariate distribution ${\mbt}_X$ and target CMR function $m_{\mbt}(x,a)$ for all $a \in \mca$ and $x \in \supp({\mbt}_X)$ is necessary and sufficient to solve problem \cref{eq:standard-goal}. In other words, the policymaker cannot obtain the best ITR $g_{\mbt}^*$ without such knowledge.\par
In this study, I consider the situation where the policymaker has no knowledge on the target conditional distribution of potential outcomes. Contrarily, I assume that they have knowledge on the target covariate distribution $\mbt_X$ and source population $\mbs \in \msp(\mcz)$, which differs from the target population. The two populations can differ in an arbitrary way as long as the following assumption is satisfied.
\begin{assumption}\label{assu:relation-source-target}
    The target population $\mbt\in \msp(\mcz)$ and source population $\mbs \in \msp(\mcz)$ satisfy the following properties:
    \begin{assuenum}
        \item\label{assuenum:covariate-absolute-continuity} The target covariate distribution ${\mbt}_X$ is absolutely continuous with respect to the source's covariate distribution ${\mbs}_X$.
        \item\label{assuenum:source-outcome-almost-bounded} The source CMR function is ${\mbs}_X$-almost surely bounded for all $a \in \mca$.
    \end{assuenum}
\end{assumption}
Similar assumptions to \cref{assuenum:covariate-absolute-continuity} have been imposed in the literature on external validity \citep[e.g.,][Assumption~3]{Hotz2005}. \cref{assuenum:source-outcome-almost-bounded} is satisfied when potential outcomes are $\mbs$-almost surely bounded. In summary, the policymaker does not know anything about the target CMR function, but instead knows the source and target covariate distribution and source CMR function. This implies that they cannot calculate the target population's welfare $V(g;\mbt)$, whatever the ITR $g$ is. Consequently, they cannot obtain the optimal ITR $g_{\mbt}^* \in \mcg$ for the target population.
\subsection{Naive Approach}\label{sec:naive-approach}
The reason the policymaker cannot obtain the optimal ITR for the target population is that the target CMR function $m_{\mbt}(x,a)$ is unknown, in contrast to the source CMR function $m_{\mbs}(x,a)$. Thus, a naive approach is to replace the target CMR function with the source CMR function and solve the optimization problem; that is, 
\begin{equation}
    g_{\text{naive-ITR}}^* \in \argmax_{g \in \mcg} \mathbf{E}_{\mbt_X}\left[\sum_{a \in \mca} m_{\mbs}(X,a)\cdot 1\{g(X) = a\}\right].
    \label{eq:naive-ITR}
\end{equation}
Note that the problem \cref{eq:naive-ITR} is well-defined as an optimization problem due to \cref{assuenum:covariate-absolute-continuity}; otherwise, there exists a point $x \in \supp(\mbt_X)$ such that $m_{\mbs}(x,a)$ is not uniquely determined. Hereafter, I refer to this approach as the \emph{naive approach} and denote its optimal solution by $g_{\text{navie-ITR}}^*$, as indicated in \cref{eq:naive-ITR}.
\par
When the target and source CMR functions are the same, the objective function in \cref{eq:naive-ITR} corresponds to the target population's welfare $V(g;\mbt)$, and hence, $g_{\text{naive-ITR}}^*$ equals $g_{\mbt}^*$. This is the case where one assumes that the source and target conditional distributions of potential outcomes are the same, as often assumed in the literature on external validity \citep[e.g., Assumption~2 of][]{Hotz2005}. Unfortunately, however, one cannot verify this assumption from the available knowledge.
\section{DR-ITR}\label{sec:DRITR}
This section formally discusses the application of DRO to the current setting. First, I define the DR-ITR with an arbitrary ambiguity set, which is a set of populations. Generally, an ambiguity set is constructed based on available knowledge. I denote the ambiguity set by $\mathscr{U}(\mbs,{\mbt}_X) \subset \mathscr{P}(\mathcal{Z})$. As a policymaker has knowledge of the source population and target covariate distribution, it is natural that the ambiguity set can depend on $\mbs$ and $\mbt_X$. Given the ambiguity set, DRO solves the following optimization problem:
\begin{equation}
    g_{\text{DR-ITR}}^{*}\in\argmax_{g\in\mathcal{G}}\underline{V}(g),
    \label{eq:DR-ITR}
\end{equation}
where 
\begin{equation}
    \underline{V}(g):=\inf_{\mbu\in\mathscr{U}(\mbs,{\mbt}_X)}V(g;\mbu).
    \label{eq:distributionally-robust-welfare}
\end{equation}
The object $\underline{V}(g)$ is the distributionally robust welfare, which is the worst-case value of welfare over the ambiguity set. Thus, DRO's goal is to choose the DR-ITR $g_{\text{DR-ITR}}^{*}$ that maximizes the distributionally robust welfare.\footnote{One of the potential drawbacks of DRO is that the max-min criterion may be too conservative to generate a valid policy. This drawback can be alleviated by adopting the Hurwicz criterion, which allows a policymaker to decide the extent to which they become conservative \citep{Hurwicz1951}. Then, the objective function can be defined as a convex combination of the worst-case and best-case values for welfare. All the results presented in this study are easily extensible with minor modifications.} Suppose that the target population $\mbt$ is contained in the ambiguity set $\mathscr{U}(\mbs,\mbt_X)$. In this case, the optimal ITR $g_{\mbt}^*$ for the target population defined in \cref{eq:standard-goal} performs better than the distributionally robust policy $g_{\text{DR-ITR}}^*$ in terms of the target population's welfare; that is, $V(g_{\mbt}^*;\mbt) \geq V(g_{\text{DR-ITR}}^*;\mbt)$. However, it is infeasible to obtain $g_{\mbt}^*$ with the current knowledge. Instead, choosing the DR-ITR ensures that the ex-post target population's welfare is at least equal to its distributionally robust welfare; that is, $V(g_{\text{DR-ITR}}^*;\mbt) \geq \underline{V}(g_{\text{DR-ITR}}^*)$.
\par
In the aforementioned DR-ITR formulation, several important issues remain unclear. First, I introduced the ambiguity set $\mathscr{U}(\mbs, \mbt_X)$ in an abstract manner. The choice of the ambiguity set is left to the policymaker's discretion, and they can construct the set based on the available information and their belief in the difference between the source and target populations. \Cref{sec:ambiguity-set-based-on-wasserstein-distance} instantiate the DR-ITR with a Wasserstein-style ambiguity set. Second, given the specific choice of the ambiguity set, the DR-ITR must evaluate the distributionally robust welfare $\underline{V}(g)$ for each ITR $g \in \mathcal{G}$. However, this is deemed difficult because the minimization problem in \cref{eq:distributionally-robust-welfare} generally involves an infinite number of distributions. Nevertheless, \cref{sec:tractable-reformulation-via-strongu-duality} shows that the distributionally robust welfare can be calculated very efficiently. In addition to these issues, I discuss what kind of population is determined as the worst case by Wasserstein DRO (\cref{sec:existence--of-worst-case-population}) and some equivalence of the DR-ITR and naive-ITR under special cases (\cref{sec:equivalence-between-dr-itr-and-naive-itr}).
\subsection{Ambiguity Set Based On Wasserstein Distance}\label{sec:ambiguity-set-based-on-wasserstein-distance}
In this study, the ambiguity set is constructed with the Wasserstein distance of order 1. Wasserstein distance is a kind of distance of probability measures that utilizes the structure of the underlying metric space. In the current setup, for a pair $(\mbs_{(Y_1,\cdots,Y_d)\mid X=x},\mbu_{(Y_1,\cdots,Y_d)\mid X=x})$ of probability measures in $\msp(\mcy^d)$, the Wasserstein distance of order 1 is defined by
\begin{equation}
    W_1(\mbs_{(Y_1,\cdots,Y_d)\mid X=x},\mbu_{(Y_1,\cdots,Y_d)\mid X=x}) := \inf_{\pi \in \Pi}\mathbf{E}_{\pi}\left[\sum_{a\in\mca}|Y_a - Y_a'|\right]
    \label{eq:wasserstein-distance}
\end{equation}
where $\Pi$ denotes the set of all the joint distributions of $(Y_1,\cdots,Y_d,Y_1',\cdots,Y_d')$, such that $(Y_1,\cdots,Y_d) \sim \mbs_{(Y_1,\cdots,Y_d)\mid X=x}$ and $(Y_1',\cdots,Y_d') \sim \mbu_{(Y_1,\cdots,Y_d)\mid X=x}$. It is well-known that there exists $\pi \in \Pi$ that attains the infimum on the right-hand side of \cref{eq:wasserstein-distance}. In addition, in our current application, it is important that the distance does not require Radon-Nikodym derivative. For a more detailed explanation, see \citet[][Chapter~6]{Villani2009}.
\par
Based on Wasserstein distance, I introduce the ambiguity set considered in this study. Concretely, given the source population $\mbs$, the target covariate distribution $\mbt_X$, and a real number $\delta \geq 0$, the ambiguity set is defined as
\begin{equation}
    \mathscr{U}(\mbs,\mbt_X) = \left\{
    \mbu \in \mathscr{P}(\mathcal{Z})
    \;\middle|\;
    \begin{multlined}
        W_1({\mbs}_{(Y_1,\cdots,Y_d)\mid X=x}, {\mbu}_{(Y_1,\cdots,Y_d)\mid X=x}) \leq \delta\\
        \text{for all $x \in \supp(\mbt_X)$ and }
        \mbu_X = \mbt_X
    \end{multlined}
    \right\}.
    \label{eq:wasserstein-ambiguity}
\end{equation}
Note that the ambiguity set \cref{eq:wasserstein-ambiguity} is well-defined under \cref{assuenum:covariate-absolute-continuity} as the source conditional distribution of potential outcomes, $\mbs_{(Y_1,\cdots,Y_d)\mid X=x}$, is unique for all $x \in \supp(\mbt_X)$. A population $\mbu$ is contained in the ambiguity set \cref{eq:wasserstein-ambiguity} only if it satisfies the following properties: (i) for each $x$ in support of the target covariate distribution $\mbt_X$, its conditional distribution of potential outcomes at point $x$ differs from the counterpart of the source population by at most $\delta$ in terms of Wasserstein distance; (ii) its marginal distribution of covariates equals the target covariate distribution, $\mbt_X$. That is, the ambiguity set point wisely constructs Wasserstein balls with the source conditional distribution of potential outcomes as the center and $\delta$ as the radius. Importantly, this allows the conditional distribution of potential outcomes to differ from that of the source population.
\par
The radius $\delta$ represents the ambiguity level of the target population, and its choice depends on the policymaker. On the one hand, the value should be large enough that the target population is contained in the ambiguity set; otherwise, it is not guaranteed that the distributionally robust welfare of the DR-ITR is a lower bound of its target population's welfare. On the other hand, the value should also be small. When the value is excessively large, the decision based on DRO can be too conservative to be helpful. Hence, the optimal value of $\delta$ would be $\delta=\sup_{x\in\supp(\mbt_{X})}W_{1}(\mbs_{(Y_{1},\cdots,Y_{d})|X=x},\mbt_{(Y_{1},\cdots,Y_{d})|X=x})$. Unfortunately, such a choice is infeasible in the current setup as knowledge on the target conditional distribution is lacking. Instead, \cref{cor:characterize-ambiguity-set} shows the relationship between certain welfare-relevant parameters and the value of $\delta$, which guides the choice of $\delta$.
\begin{corollary}\label{cor:characterize-ambiguity-set}
    Suppose that \cref{assu:relation-source-target} is in force, and assume that a population $\mbu \in \mathscr{P}(\mathcal{Z})$ is in the ambiguity set $\mathscr{U}(\mbs, \mbt_X)$ given in \cref{eq:wasserstein-ambiguity}. Then, the population $\mbu$ satisfies the following inequalities:
    \begin{corenum}[label=(\roman*),ref=\thecorollary.(\roman*)]
        \item \label{corenum:charact-cmr}For all $a\in\mca$ and $x \in \supp(\mbt_X)$,
        \begin{equation*}
            |m_{\mbu}(x,a)-m_{\mbs}(x,a)|\leq\delta.
        \end{equation*}
        \item \label{corenum:charact-cate}For all $a_1,a_2\in\mca$ and $x \in \supp(\mbt_X)$,
        \begin{equation*}
            |(m_{\mbu}(x,a_1)-m_{\mbu}(x,a_2))-(m_{\mbs}(x,a_1)-m_{\mbs}(x,a_2))|\leq\delta.
        \end{equation*}
        \item \label{corenum:charact-amr}For all $a\in\mca$,
        \begin{equation*}
            |\mathbf{E}_{\mbu}[Y_{a}]-\mathbf{E}_{\mbs}[Y_{a}]|\leq\delta+|\mathbf{E}_{\mbt_{X}}[m_{\mbs}(X,a)]-\mathbf{E}_{\mbs_{X}}[m_{\mbs}(X,a)]|.
        \end{equation*}
        \item \label{corenum:charact-ate} For all $a_1,a_2\in\mca$,
        \begin{align*}
            |\mathbf{E}_{\mbu}[Y_{a_1}-Y_{a_2}]-\mathbf{E}_{\mbs}[Y_{a_1}-Y_{a_2}]|
            \leq\delta+|\mathbf{E}_{\mbt_{X}}[m_{\mbs}(X,a_1)-m_{\mbs}(X,a_2)]-\mathbf{E}_{\mbs_{X}}[m_{\mbs}(X,a_1)-m_{\mbs}(X,a_2)]|.
        \end{align*}
    \end{corenum}
\end{corollary}
The first two items characterize the difference in the conditional distributions of potential outcomes for each possible $x$. In particular, the first one states that as long as a population is in the ambiguity set, its CMR function differs from the source CMR function by at most $\delta$. Similarly, the second one states that the difference in the CMR functions between any pairs of treatments differs from the counterpart of the source population by at most $\delta$. The latter two items characterize the difference in the marginal distributions of potential outcomes, while focusing on its mean. For the third item, the upper bound consists of the ambiguity level $\delta$ and an additional term that essentially measures the difference in the covariate distributions $\mbt_X$ and $\mbs_X$. A similar interpretation holds for the fourth item. Note that this result does not address the tightness of the aforementioned inequalities. In particular, the result does not imply the existence of a distribution $\mbu\in\mathscr{U}(\mbs,\mbt_X)$ with $|m_{\mbu}(x,a)-m_{\mbs}(x,a)|=\delta$ for some $x$ and $a$. The existence of such distributions are guaranteed in \cref{cor:existence-of-worst-case-population}, which will be presented later. 
\begin{remark}[Ambigutiy set based on the Wasserstein distance of general order]
    One can construct the ambiguity set using the Wasserstein distance of general order. Specifically, for $p \in [1,\infty)$, the Wasserstein distance of order $p$ is defined by
    \begin{equation*}
        W_p(\mbs_{(Y_1,\cdots,Y_d)\mid X=x},\mbu_{(Y_1,\cdots,Y_d)\mid X=x}) := \inf_{\pi \in \Pi} \left(\mathbf{E}_\pi\left[\left(\sum_{a \in \mca}|Y_a - Y_a'|\right)^p\right]\right)^{1/p}
    \end{equation*}
    for any pair $(\mbs_{(Y_1,\cdots,Y_d)\mid X=x},\mbu_{(Y_1,\cdots,Y_d)\mid X=x}) \in \msp(\mcy^d)\times\msp(\mcy^d)$.
\end{remark}
\begin{remark}[Ambiguity set based on $\phi$-divergence]
    For a convex function $\phi:[0,\infty)\to(-\infty,\infty]$, the $\phi$-divergence between distributions $\mbs_{(Y_1,\cdots,Y_d)\mid X=x}$ and $\mbu_{(Y_1,\cdots,Y_d)\mid X=x}$ with $\mbu_{(Y_1,\cdots,Y_d)\mid X=x} \ll \mbs_{(Y_1,\cdots,Y_d)\mid X=x}$ is defined by
    \begin{equation*}
        D_{\phi}(\mbu_{(Y_1,\cdots,Y_d)\mid X=x}\Vert\mbs_{(Y_1,\cdots,Y_d)\mid X=x}) := \int \phi\left(\frac{\mathrm{d}\mbu_{(Y_1,\cdots,Y_d)\mid X=x}}{\mathrm{d}\mbs_{(Y_1,\cdots,Y_d)\mid X=x}}\right) d\mbs_{(Y_1,\cdots,Y_d)\mid X=x}.
    \end{equation*}
    This definition includes the Kullback-Leibler (KL) divergence, Hellinger distance, or $\chi^2$-divergence as a special case. One can construct $\phi$-divergence based ambiguity set in \cref{eq:wasserstein-ambiguity} by replacing $W_1$ with $D_{\phi}$. The crucial advantage of Wasserstein over $\phi$-divergence is that a ball based on $\phi$-divergence requires the support of each element in the ball to be included in the support of the central distribution, whereas the ball based on Wasserstein distance does not. Such requirements highly restrict the applicability of DRO.
\end{remark}

\subsection{Tractable Reformulation via Strong Duality}\label{sec:tractable-reformulation-via-strongu-duality}
To solve problem \cref{eq:DR-ITR}, one must evaluate the distributionally robust welfare $\underline{V}(g)$ in \cref{eq:distributionally-robust-welfare}. However, the infimum generally involves an infinite number of distributions and is not easy to calculate directly. In this section, I provide a tractable reformulation of the problem \cref{eq:DR-ITR} by exploiting the strong duality results from \citet{Blanchet2019a}, which simplifies our analysis.
\par
First, the law of iterated expectations implies that the minimization problem \cref{eq:distributionally-robust-welfare} can be solved component-wise. Specifically, it can be written as $\underline{V}(g) = \mathbf{E}_{\mbt_X}[\underline{v}(X;g)]$, where
\begin{equation}
    \underline{v}(x;g) := 
    \inf_{\substack{\mbu_{(Y_{1},\cdots,Y_{d})|X=x} \in \mathscr{P}(\mathcal{Y}^d) \\ \text{s.t.  }W_1(\mbs_{(Y_{1},\cdots,Y_{d})|X=x},\mbu_{(Y_{1},\cdots,Y_{d})|X=x}) \leq \delta}}\mathbf{E}_{\mbu_{(Y_{1},\cdots,Y_{d})|X=x}}\left[\sum_{a\in\mathcal{A}}Y_{a}\cdot1\{g(x)=a\}\right]
    \label{eq:worst-case-cond-welfare}
\end{equation}
for each $x \in \supp(\mbt_X)$. The object $\underline{v}(x;g)$ can be interpreted as the worst-case conditional welfare of policy $g$ at covariate value $x$. With these notations, the evaluation must be reduced to $\underline{v}(x;g)$ for all $x\in\supp(\mbt_{X})$ and $g \in \mathcal{G}$.
\par
One can obtain the strong dual problem of \cref{eq:worst-case-cond-welfare} by applying the results of \citet{Blanchet2019a}. In \ifdefined\OVERALL\cref{sec:review-of-BM}\else\cref{appendix-sec:review-of-BM} in the supplementary materials\fi, I briefly review their result. The significant advantage of this result in our application is that the dual problem involves only the source conditional distribution and that the dual problem involves a one-dimensional concave programming with respect to the dual argument. By explicitly solving this dual problem, the following theorem is obtained.
\begin{theorem}\label{thm:tractable-reformulation-dr-itr}
    Suppose that \cref{assu:relation-source-target} is in force. Then, for each $x \in \supp(\mbt_X)$ and $g \in \mathcal{G}$, it holds that
    \begin{equation*}
        \underline{v}(x;g)=\sum_{a\in\mathcal{A}}\max\{m_{\mbs}(x,a)-\delta,\inf\mathcal{Y}\}\cdot1\{g(x)=a\},
    \end{equation*}
    where $\inf\mcy$ is the infimum of outcome space $\mcy$, and it is set to $-\infty$ when $\mcy$ is unbounded from below. Therefore, the DR-ITR can be equivalently represented by
    \begin{equation}
        g_{\text{DR-ITR}}^* \in \argmax_{g \in \mathcal{G}} \mathbf{E}_{\mbt_X}\left[\sum_{a \in \mathcal{A}}\max\{m_{\mbs}(x,a)-\delta,\inf\mathcal{Y}\}\cdot 1\{g(X) = a\}\right].
        \label{eq:tractable-dr-itr}
    \end{equation}
\end{theorem}
As is clear from the first equation of \cref{thm:tractable-reformulation-dr-itr}, the worst-case conditional welfare can be written in a simpler form, and its interpretation becomes simple as well. Consider an ITR $g \in \mathcal{G}$ that assigns treatment $a \in \mathcal{A}$ to individuals with the covariate $X=x$. The worst-case conditional welfare $\underline{v}(x;g)$ of the ITR for these individuals is obtained by translating the source CMR function of treatment $a$ by $-\delta$ (when such a translation is unrealistic owing to the lower bound of the outcome space, the worst-case value is set at the lower bound). Because of this simplification, the DR-ITR becomes tidy, as shown in \cref{eq:tractable-dr-itr}. This result provides an interesting connection with the naive approach in \cref{sec:equivalence-between-dr-itr-and-naive-itr}, and also underlies the estimation method discussed in \cref{sec:estimation}. Note that the above result does not address the existence of a worst-case population that attains the infimum in \cref{eq:distributionally-robust-welfare}, which will be discussed in \cref{sec:existence--of-worst-case-population}.
\begin{remark}\label{remark:KL-ambiguity-duality}
    Even when the ambiguity set is specified using the Kullback-Leibler divergence instead of the 1-Wasserstein distance, the worst-case conditional welfare can be simplified using its strong dual representation. Specifically, the strong duality result from \cite[][Theorem~1 and Theorem~2]{Hu2012} implies that the worst-case conditional welfare of the ITR $g\in\mathcal{G}$ at $x\in\mathcal{X}$ can be obtained as the optimal value for the following problem:
    \begin{equation*}
        \max_{\lambda\geq0}\left\{-\lambda\delta-\lambda\log\mathbf{E}_{\mbs_{(Y_1,\cdots,Y_d)\mid X=x}}\left[\exp\left(\frac{-\sum_{a\in\mathcal{A}}Y_a\cdot1\{g(x)=a\}}{\lambda}\right)\right]\right\}.
    \end{equation*}
    However, as opposed to the case of the 1-Wasserstein distance, the optimization problem cannot be analytically solved for the general source population. Therefore, such a concise representation of the DR-ITR as \cref{eq:tractable-dr-itr} is not available for the KL divergence-based ambiguity set. This will introduce computational complexity when considering the estimation procedure.
\end{remark}
\subsection{Existence of the Worst-Case Population}\label{sec:existence--of-worst-case-population}
It is natural to question whether there exists a worst-case population that attains the infimum in \cref{eq:distributionally-robust-welfare}, and if so, what kind of population it is. This section discusses the existence and characterization of such populations in the DR-ITR for specific outcome spaces. Again, the application of the results of \citet{Blanchet2019a} helps determine the existence and characterization of the worst-case population, as shown in \cref{cor:existence-of-worst-case-population}.
\begin{corollary}\label{cor:existence-of-worst-case-population}
    Suppose that $\delta > 0$, $x \in \supp(\mbt_X)$.
    \begin{enumerate}[label=(\roman*),ref=\thecorollary.(\roman*)]
        \item If $\mathcal{Y}$ is convex and unbounded from below, the worst-case conditional distribution, which attains the infimum in \cref{eq:worst-case-cond-welfare}, exists.
        Moreover, one of such distributions is characterized as an induced measure $t_{\#}\mbs_{(Y_{1},\cdots,Y_{d})\mid X=x}$, with 
        \begin{equation*}
            t(y_{1},\cdots,y_{d})=(y_{1},\cdots,y_{g(x)-1},y_{g(x)}-\delta,y_{g(x)+1},\cdots,y_{d}).
        \end{equation*}
        \item If $\mathcal{Y}$ is convex and bounded from below, the worst-case conditional distribution, which attains the infimum in \cref{eq:worst-case-cond-welfare}, exists. Moreover, one of such distributions is characterized as an induced measure $t_{\#}\mbs_{(Y_{1},\cdots,Y_{d})\mid X=x}$, with 
        \begin{equation*}
            t(y_{1},\cdots,y_{d})=
            \begin{cases}
                \left(y_{1},\cdots,y_{g(x)-1},y_{g(x)}-\frac{\delta(y_{g(x)}-\inf\mathcal{Y})}{m_{\mbs}(x,g(x))-\inf\mathcal{Y}},y_{g(x)+1},\cdots,y_{d}\right) & \text{if } m_{\mbs}(x,g(x))-\delta>\inf\mathcal{Y}\\
                (y_{1},\cdots,y_{g(x)-1},\inf\mathcal{Y},y_{g(x)+1},\cdots,y_{d}) & \text{if } m_{\mbs}(x,g(x))-\delta\leq\inf\mathcal{Y}
            \end{cases}
        \end{equation*}
    \end{enumerate}
\end{corollary}
\cref{cor:existence-of-worst-case-population} shows the existence of the worst-case conditional distribution of potential outcomes, which attains the infimum in \eqref{eq:worst-case-cond-welfare} when the outcome space $\mathcal{Y}$ is a convex subset of $\mathbb{R}$. The result indicates that the worst-case conditional distribution of potential outcomes always exists when the outcome space is convex. Then, one can obtain the worst-case population by incorporating the worst-case conditional distribution with the target covariate distribution $\mbt_X$.
\par
The corollary also gives examples of the worst-case conditional distribution. When the outcome space is convex and unbounded from below, one worst-case conditional distribution can be obtained by simply moving the source conditional distribution $\mbs_{(Y_1,\cdots,Y_d)\mid X=x}$ by $-\delta$ along the $g(x)$-th dimension. Consequently, the worst-case conditional welfare is equal to $m_{\mbs}(x,g(x)) - \delta$. When the outcome space is convex but bounded from below, the characterization depends on whether it is logically possible to decrease the source CMR by $\delta$. When $m_{\mbs}(x,g(x))-\delta > \inf\mcy$, the worst-case distribution can be obtained by reducing the source CMR by $\delta$ and concentrating more mass around the mean. Conversely, if $m_{\mbs}(x,g(x))-\delta\leq\inf\mathcal{Y}$, the worst-case conditional distribution concentrates on $\inf\mathcal{Y}$ for the $g(x)$-th dimension. Note that the distributions given in the corollary are merely examples of worst-case conditional distributions. For example, when $\mathcal{Y}=\mathbb{R}$, the worst-case distribution can also be characterized as an induced measure using a map
\begin{equation*}
    t(y_{1},\cdots,y_{d})=\left(y_{1},\cdots,y_{g(x)-1},y_{g(x)}-\frac{\delta}{m_{\mbs}(x,g(x))}y_{g(x)},y_{g(x)+1},\cdots,y_{d}\right)
\end{equation*}
given $m_{\mbs}(x,g(x))\neq0$. Finally, this result implies that the bound given in \cref{corenum:charact-cmr} is tight.
\subsection{Relation Between the DR-ITR and Naive-ITR}\label{sec:equivalence-between-dr-itr-and-naive-itr}
This subsection discusses an interesting relationship between the DR-ITR and naive-ITR in \cref{sec:naive-approach}, which provides a justification for the naive approach from the perspective of Wasserstein DRO. Let $g_{\mbs}^{\text{FB}}:\mcx \to \mca$ be the first best ITR for the source population such that $g_{\mbs}^{\text{FB}}(x) \in \argmax_{a \in \mca} m_{\mbs}(x,a)$ $\mbs_X$-a.s.. I additionally impose the following assumption.
\begin{assumption}\label{assu:equivalence-between-dr-naive}
    Either of the following is true:
    \begin{assuenum}
        \item\label{assuenum:source-fb-contained} The first best ITR $g_{\mbs}^{\text{FB}}$ for the source population is an element of $\mcg$.
        \item\label{assuenum:source-cmr-distant-from-bound} For any policy $g\in\mathcal{G}$, it holds $\mbt_X$-almost surely that $m_{\mbs}(x,g(x)) - \delta \geq \inf\mathcal{Y}$.
    \end{assuenum}
\end{assumption}
A simple example in which \cref{assuenum:source-fb-contained} is satisfied is the case where $\mcg$ consists of all measurable mappings from $\mcx$ to $\mca$.
\Cref{assuenum:source-cmr-distant-from-bound} informally states that the source CMR function $m_{\mbs}(x,a)$ is sufficiently distant from the lower bound of the outcome space. One typical case is where the outcome space $\mathcal{Y}$ is unbounded from below; that is, $\inf\mathcal{Y}=-\infty$. In this case, the assumption holds regardless of the value of $\delta$ and the class $\mathcal{G}$ of the ITRs. Under \cref{assu:equivalence-between-dr-naive}, one has the following theorem:
\begin{theorem}\label{thm:equiv-between-dr-naive}
    Suppose that \cref{assu:relation-source-target,assu:equivalence-between-dr-naive} hold. Then, any naive-ITR is also a DR-ITR.
\end{theorem}
At first glance, this result seems unnatural. The naive approach imposes a somewhat strong assumption so that the target and source CMR functions become identical, and then searches for the best ITR. Contrarily, the DR-ITR considers populations with CMR functions that can deviate from the source CMR function more flexibly and optimizes against the worst-case over such populations. Hence, there appears to be a difference between the ITRs resulting from these two approaches. However, when the ambiguity set is specified using the 1-Wasserstein distance as in \cref{eq:wasserstein-ambiguity} and \cref{assu:equivalence-between-dr-naive} holds, the naive-ITR is already distributionally robust. 
\par
I explain the relation between \cref{assu:equivalence-between-dr-naive,thm:equiv-between-dr-naive} in more depth. For \cref{assuenum:source-fb-contained}, notice that \cref{thm:tractable-reformulation-dr-itr} implies that the worst-case conditional welfare $\underline{v}(x;g)$ is always maximized by choosing the first best ITR $g_{\mbs}^{\text{FB}}$. Thus, $g_{\mbs}^{\text{FB}}$ also maximizes the worst-case welfare $\underline{V}(g)$. Therefore, as long as the first best ITR is in $\mcg$, it is one of the DR-ITRs. However, when \cref{assuenum:source-cmr-distant-from-bound} holds, \cref{thm:tractable-reformulation-dr-itr} implies that 
\begin{equation*}
    \underline{V}(g) = \mathbf{E}_{\mbt_X}\left[\sum_{a \in \mca}m_{\mbs}(X,a)\cdot 1\{g(X) = a\}\right] - \delta
\end{equation*}
for all $g \in \mcg$. That is, the worst-case welfare is obtained by translating the objective function of the naive approach by $-\delta$. Thus, maximizing the distributionally robust welfare is equivalent to maximizing the objective function of the naive approach.
\par
Finally, I note that this consequence does not hold when \cref{assu:equivalence-between-dr-naive} is not satisfied as exemplified in \cref{ex:counterexample-equiv-between-naive-and-drplw}. In addition, such a relation does not necessarily hold when the ambiguity set is constructed using other metrics as demonstrated in \cref{remark:KL-ambiguity-duality}.
\begin{example}\label{ex:counterexample-equiv-between-naive-and-drplw}
    Suppose that $\mathcal{Y} = [0,\infty)$, $\mathcal{A} = \{0,1\}$, $\mathcal{X} = \{m,f\}$, $\delta = 1$, and $\mathcal{G} = \{g_1,g_2\}$ such that
    \begin{equation*}
        g_1(x) = 
        \begin{cases}
            1 & \text{if } x = m\\
            0 & \text{if } x = f
        \end{cases}
        \quad\text{and}\quad
        g_2(x) = 
        \begin{cases}
            0 & \text{if } x = m\\
            1 & \text{if } x = f.
        \end{cases}
    \end{equation*}
    In addition, assume that the source CMR function and target covariate distribution are given as follows:
    \begin{equation*}
        m_{\mbs}(x,a) = 
        \begin{cases}
            0.5 & \text{if } x = m\text{ and }a=1\\
            0.4 & \text{if } x = m\text{ and }a=0\\
            1.5 & \text{if } x = f\text{ and }a=1\\
            1.4 & \text{if } x = f\text{ and }a=0
        \end{cases}
        \quad\text{and}\quad
        \mbt_{X}(x) = 
        \begin{cases}
            q & \text{if } x = m,\\
            1-q & \text{if } x = f,
        \end{cases}
    \end{equation*}
    where $q \in (0.5,1)$. Observe that \cref{assu:equivalence-between-dr-naive} does not hold in this setting; in fact, one has $m_{\mbs}(m,g_1(m)) - \delta = -0.5$, although $\mbt_X(\{m\}) > 0$. It is easily shown that the naive approach chooses $g_1$, while DRO chooses $g_2$.
\end{example}
\begin{remark}
    When the ambiguity set is specified using the Kullback-Leibler divergence, the naive-ITR and DR-ITR generally do not agree. Consider the case in which $\mathcal{Y} = \mathbb{R}$ and $\mbs_{Y_a\mid X=x}=N(m_{\mbs}(x,a),\sigma_{\mbs}^2(x,a))$ for each $a \in \mathcal{A}$ and $x \in \supp(\mbt_X)$. In this case, $\exp(-Y_{g(x)}/\lambda)$ is distributed according to a log-normal distribution with the parameters $-m_{\mbs}(x,g(x))/\lambda$ and $\sigma_{\mbs}(x,g(x))/\lambda$. Hence, its expectation is represented by $\exp(-m_{\mbs}(x,g(x))/\lambda + \sigma_{\mbs}^2(x,g(x))/(2\lambda^2))$. Based on the strong duality result discussed in \cref{remark:KL-ambiguity-duality}, DRO with the Kullback-Leibler based ambiguity set maximizes the following function over the policy class:
    \begin{equation*}
        \mathbf{E}_{\mbt_X}\left[\sum_{a\in\mathcal{A}}\left(m_{\mbs}(X,a)-\sqrt{2\delta}\sigma_{\mbs}(X,a)\right)\cdot 1\{g(X)=a\}\right].
    \end{equation*}
    It is evident that the objective function differs from that of the naive approach; hence, the solutions of these two approaches are not similar.
\end{remark}
\subsection{Consideration Regarding the Unknown Covariate Shift}
Thus far, it has been assumed that the target covariate distribution is known. However, what if it is unknown? Even in such a case, one can construct another ambiguity set on the target covariate distribution, and seek a DR-ITR that is distributionally robust against the unknown covariate shifts as well. In particular, if \cref{assuenum:covariate-absolute-continuity} is satisfied, the ambiguity set \citep[e.g.,][]{Mo2021,Zhao2019} can be utilized.
\par
Here, I incorporate the above idea with the ambiguity set based on $\phi$-divergence. Suppose that a policymaker has no knowledge about the target population, but instead has knowledge on the source population. In addition, I assume that \cref{assu:relation-source-target} holds. Then, one can incorporate unknown covariate shifts by using the following ambiguity set:
\begin{equation*}
    \mathscr{U}(\mbs)= \left\{
    \mbu \in \mathscr{P}(\mathcal{Z})
    \;\middle|\;
    \begin{multlined}
        W_1({\mbs}_{(Y_1,\cdots,Y_d)\mid X=x}, {\mbu}_{(Y_1,\cdots,Y_d)\mid X=x}) \leq \delta\\
        \text{for all $x \in \supp(\mbu_X)$ and }
        D_{\phi}(\mbu_X \Vert \mbs_X) \leq \rho
    \end{multlined}
    \right\}
\end{equation*}
for $\rho\geq 0$. With this ambiguity set, the DR-ITR is an ITR that maximizes the worst-case welfare over the ambiguity set; that is, $g_{\text{DR-ITR}}^* \in \argmax_{g \in \mcg} \inf_{\mbu \in \mathscr{U}(\mbs)} V(g;\mbu)$. As the constraints on the conditional distribution of potential outcomes and the covariate distribution are independent, one can calculate the distributionally robust welfare in a two-step procedure: (i) obtain the worst-case conditional welfare $\underline{v}(x;g)$ and (ii) calculate the worst-case value of the marginalized worst-case conditional welfare over the ambiguity set. Thus, \cref{thm:tractable-reformulation-dr-itr} implies that the distributionally robust welfare is expressed as
\begin{equation*}
    \inf_{\substack{\mbu_X \in \msp(\mcx) \\\text{ s.t. } D_{\phi}(\mbu_X\Vert\mbs_X) \leq \rho}} \mathbf{E}_{\mbu_X}\left[\sum_{a \in \mca} \max\{m_{\mbs}(X,a)-\delta,\inf\mcy\}\cdot 1\{g(X) = a\}\right].
\end{equation*}
\par
As in \cref{sec:equivalence-between-dr-itr-and-naive-itr}, \cref{assu:equivalence-between-dr-naive} provides an interesting result. Specifically, it can be shown that under the assumption it holds that
\begin{equation*}
    \begin{aligned}
        &\argmax_{g \in \mcg} \inf_{\substack{\mbu_X \in \msp(\mcx) \\\text{ s.t. } D_{\phi}(\mbu_X\Vert\mbs_X) \leq \rho}} \mathbf{E}_{\mbu_X}\left[\sum_{a \in \mca} m_{\mbs}(X,a)\cdot 1\{g(X) = a\}\right]\\
        \subset & \argmax_{g \in \mcg} \inf_{\substack{\mbu_X \in \msp(\mcx) \\\text{ s.t. } D_{\phi}(\mbu_X\Vert\mbs_X) \leq \rho}} \mathbf{E}_{\mbu_X}\left[\sum_{a \in \mca} \max\{m_{\mbs}(X,a)-\delta,\inf\mcy\}\cdot 1\{g(X) = a\}\right].
    \end{aligned}
\end{equation*}
Thus, the DR-ITRs obtained by ignoring the possible difference in conditional distributions of potential outcomes are also distributionally robust. This result specifically reinforces the DR-ITR considered in \citet{Mo2021}.
\section{Estimation}\label{sec:estimation}
In this section, I discuss the identification and estimation of the DR-ITR $g_{\text{DR-ITR}}^{*}$. Then, I evaluate the theoretical performance of its estimator. The estimation method is based on \cref{thm:tractable-reformulation-dr-itr}.
\par
Suppose that the experimental or observational data generated from the source population $\mbs$ and the covariate data generated from the target population $\mbt$ are available. Specifically, the data from the source population are $\{(Y_{i},A_{i},X_{i}^{s})\}_{i=1}^{n_{s}}$, where $X_{i}^{s}\in\mathcal{X}$ refers to the observable pre-treatment covariates of individual $i$, $A_{i}\in\mathcal{A}$ denotes the individual's treatment assignment, and $Y_{i}\in\mathcal{Y}$ is the post-treatment observed outcome. The sample is an i.i.d. draw from a joint distribution of $(Y_{1,i},\cdots,Y_{d,i},A_i,X_i)$. I assume that it satisfies unconfoundedness, the overlap condition, and consistency, and that its marginal distribution of $(Y_{1,i},\cdots,Y_{d,i},X_i)$ is the same as the source population $\mbs$. However, the covariate data from the target population is $\{X_{j}^{t}\}_{j=1}^{n_{t}}$, where each $X_{j}^{t}$ is an i.i.d. draw from $\mbt_{X}$.
\par
The following are identifiable from this data; the marginal distributions $\mbs_{Y_a\mid X=x}$ of the source conditional distribution of potential outcomes for each $a \in \mca$ and $x \in \supp(\mbs_X)$, the source covariate distribution $\mbs_X$, and the target covariate distribution $\mbt_X$. Thus, it is obvious from \cref{thm:tractable-reformulation-dr-itr} that the distributionally robust welfare is identifiable, and thus, a natural estimator for $g_{\text{DR-ITR}}^*$ can be constructed as follows:
\begin{enumerate}[label=(\roman*)]
    \item Using data $\{(Y_{i},A_{i},X_{i}^{s})\}_{i=1}^{n_{s}}$, estimate $m_{\mbs}(x,a)$ to obtain $\widehat{m}_{\mbs}(x,a)$ for all $a \in \mca$ and $x \in \supp(\mbs_X)$.
    \item Solve 
    \begin{equation}
        \widehat{g}_{\text{DR-ITR}}\in\argmax_{g\in\mathcal{G}}\widehat{\underline{V}}(g)
        \label{eq:drpl-estimator}
    \end{equation}
    where 
    \begin{equation}
        \widehat{\underline{V}}(g):=\frac{1}{n_{t}}\sum_{j=1}^{n_{t}}\sum_{a\in\mathcal{A}}\max\{\widehat{m}_{\mbs}(X_{j}^{t},a)-\delta,\inf\mathcal{Y}\}\cdot1\{g(X_{j}^{t})=a\}.
        \label{eq:estimator-worst-case-welfare}
    \end{equation}
\end{enumerate}
The object $\widehat{\underline{V}}(g)$ is an estimator of the distributionally robust welfare. The estimator is constructed by replacing the unknown source CMR function $m_{\mbs}(x,a)$ with the estimated CMR function $\widehat{m}_{\mbs}(x,a)$, and replacing the unknown target covariate distribution $\mbt_X$ with the empirical distribution based on $\{X_j^t\}_{j=1}^t$. Arbitrary method can be used as the estimator for the source CMR function.
\begin{remark}
    As demonstrated in \cref{sec:equivalence-between-dr-itr-and-naive-itr}, when one assumes \cref{assu:equivalence-between-dr-naive}, it suffices to seek a naive-ITR, to obtain a DR-ITR. Hence, the problem boils down to the question of how to estimate the objective function in \cref{eq:naive-ITR}. This problem has already been analyzed in existing studies, for example, \citet[Remark~2.2]{Kitagawa2018} and \citet{Uehara2020}.
\end{remark}
\subsection{Theoretical Guarantee}\label{sec:theoretical-guarantee}
In line with the literature on policy learning, I evaluate the estimator's performance using a kind of regret. Usually, the regret of the ITR $g$ is defined as the difference in the target population's welfare between the optimal ITR $g_{\mbt}^*$ for the target population and the ITR $g$; that is, the usual regret is defined as $V(g_{\mbt}^*;\mbt)-V(g;\mbt)$. However, the goal of DRO is different from that of standard policy learning; therefore, it is natural to modify the definition of regret. Specifically, I focus on the \emph{distributionally robust regret} $R_{\text{DRO}}(g)$ of the ITR $g$ defined by 
\begin{equation*}
    R_{\text{DRO}}(g):=\underline{V}(g_{\text{DR-ITR}}^*)-\underline{V}(g).
\end{equation*}
In contrast to the usual regret, distributionally robust regret is defined as the difference in the distributionally robust welfare between the true DR-ITR and the ITR $g$. The same concept is also considered by, for example, \citet{Lee2018, Si2021, Tu2019}. In the following, I derive a high probability bound on $R_{\mathrm{DRO}}(\widehat{g}_{\text{DR-ITR}})$. 
\par
I begin by listing some assumptions necessary for the analysis. Thereafter, I denote the expectation with respect to the distribution of data by $\mathbf{E}[\cdot]$.
\begin{assumption}\label{assu:dgp}
    The data generating process satisfies the following conditions:
    \begin{assuenum}
        \item $\{(Y_{i},A_{i},X_{i}^{s})\}_{i=1}^{n_{s}}\perp\{X_{j}^{t}\}_{j=1}^{n_{t}}$.\label{assu-item:independence}
        \item One has access to an estimator $\widehat{m}_{\mbs}(x,a)$ for $m_{\mbs}(x,a)$ such that $\widehat{m}_{\mbs}(x,a)$ are bounded almost surely and 
        \begin{equation*}
            C_{1}:=\limsup_{n_{s}\to\infty}\psi_{n_{s}}\mathbf{E}\left[\sum_{a\in\mathcal{A}}\left|\widehat{m}_{\mbs}(X,a)-m_{\mbs}(X,a)\right|^{2}\right]<\infty
        \end{equation*}
         for $\psi_{n_{s}}\to\infty$ as $n_{s}\to\infty$.\label{assu-item:convergence-rate-of-mse}
    \end{assuenum}
\end{assumption}
\begin{assumption}\label{assu:policy-class}
    There exist constants $C > 0$, $D > 0$, and $0 < w < \frac{1}{2}$ such that $N_H(u,\mcg) \leq C \exp(D/u^w)$ for all $u \in (0,1)$.
\end{assumption}
\cref{assu:dgp} is the assumption on the data generating process. \cref{assu:policy-class} is included to control the complexity of the ITR class. The object $N_H(u,\mcg)$ in the assumption is the covering number of $\mcg$, with $u$ as the radius and the Hamming distance as the distance. For its precise definition, see Definition~4 of \citet{Zhou2022}. Then, under \cref{assu:policy-class}, I introduce the entropy integral $\kappa(\mcg)$ by $\kappa(\mathcal{G}):=\int_{0}^{1}\sqrt{\log N_{H}(u^{2},\mathcal{G})}du < \infty$. Combining these assumptions, the next theorem follows.
\begin{theorem}\label{thm:performance-guarantee}
    Suppose that \cref{assu:relation-source-target,assu:dgp,assu:policy-class} are satisfied. Then, for any $\epsilon\in(0,1)$, with a probability of at least $1-\epsilon$, it holds that 
    \begin{equation*}
        R_{\text{DRO}}(\widehat{g}_{\text{DR-ITR}})\leq\left(54.4\sqrt{2}(\kappa(\mathcal{G})+8)+\sqrt{2\log\frac{2}{\epsilon}}\right)\sqrt{\frac{2d(M+\delta)^{2}}{n_{t}}}+\sqrt{\frac{8C_{1}}{\epsilon\psi_{n_{s}}}}+o\left(\frac{1}{\sqrt{n_{t}}}\right),
    \end{equation*}
    where $M<\infty$ is an almost sure bound of $m_{\mbs}(x,a)$.
\end{theorem}
\cref{thm:performance-guarantee} provides a high-probability bound on the distributionally robust regret. The first term of the upper bound is a high probability bound on the error generated using the empirical distribution of $\{X_{j}^{t}\}_{j=1}^{n_{t}}$, instead of the true target covariate distribution $\mbt_{X}$. Contrastingly, the second term is a high probability bound on the error generated by using the estimator $\widehat{m}_{\mbs}(x,a)$, instead of the true source CMR function $m_{\mbs}(x,a)$. Overall, the distributionally robust regret with high probability converges to zero at a rate of $O(n_{t}^{-1/2}\vee\psi_{n_{s}}^{-1/2})$.

\section{Empirical Application}\label{sec:empirical-application}
I demonstrate the proposed DR-ITR using data from the experimental evaluations of changes in welfare-to-work programs during the early 1980s. The program changes were aimed to increase employment and decrease welfare receipt. The new programs required single-parent families who were receiving benefits from Aid to Families with Dependent Children (AFDC), to participate in employment support programs as a condition for receiving the full amount of their monthly welfare payments. For a more detailed explanation of the program changes, see \citet{Friedlander1995}, \citet{Hotz2005}, and the references therein. Several states conducted randomized controlled trials (RCTs) to evaluate the effects of the program changes. Among these states, I focus on two experiments conducted in Arkansas and San Diego, because these two states seem to be very different in terms of the covariate distribution and conditional distribution of potential outcomes.
\par
The new programs implemented and the corresponding experiments in Arkansas and San Diego differed slightly in terms of timing, central target family, and program components. The Arkansas WORK Program (WORK) targeted AFDC applicants and recipients with children aged at least three years and provided group job search and unpaid work experience. The corresponding experiment started in 1983 with 1,127 participants, of whom half were exposed to WORK and the rest were assigned to the control condition. Contrarily, the Saturation Work Initiative Model (SWIM) in San Diego targeted AFDC applicants and recipients with children aged at least six years and provided job search assistance, skills training, and unpaid work experience. The corresponding experiment started in 1985 with 3,211 participants, of whom half were exposed to the SWIM, and the remainder were assigned to the control condition. In both RCTs, the covariates related to personal characteristics and employment information were collected prior to the experiments. In addition, both experiments involved following up with participants at least two years after the start of the experiments, and collected earnings as outcome variables.
\par
For the illustration of the DR-ITR, I consider the following hypothetical situation: Suppose that the government of Arkansas desires a good ITR that specifies whether an individual should participate in the WORK or not. Assume that the data available to the government are limited only to the data from RCT on the SWIM and the data of covariates from RCT on the WORK. In other words, I assume that the experimental assignment variable and the outcome variable of the WORK are not observed. Thus, according to the terminology of this study, the target population comprises single-parent families in Arkansas, whereas the source population comprises single-parent families in San Diego.
\par
\begin{figure}[t]
    \centering
    \caption{Difference in the conditional average treatment effect between Arkansas and San Diego}
    \begin{threeparttable}
    \includegraphics[]{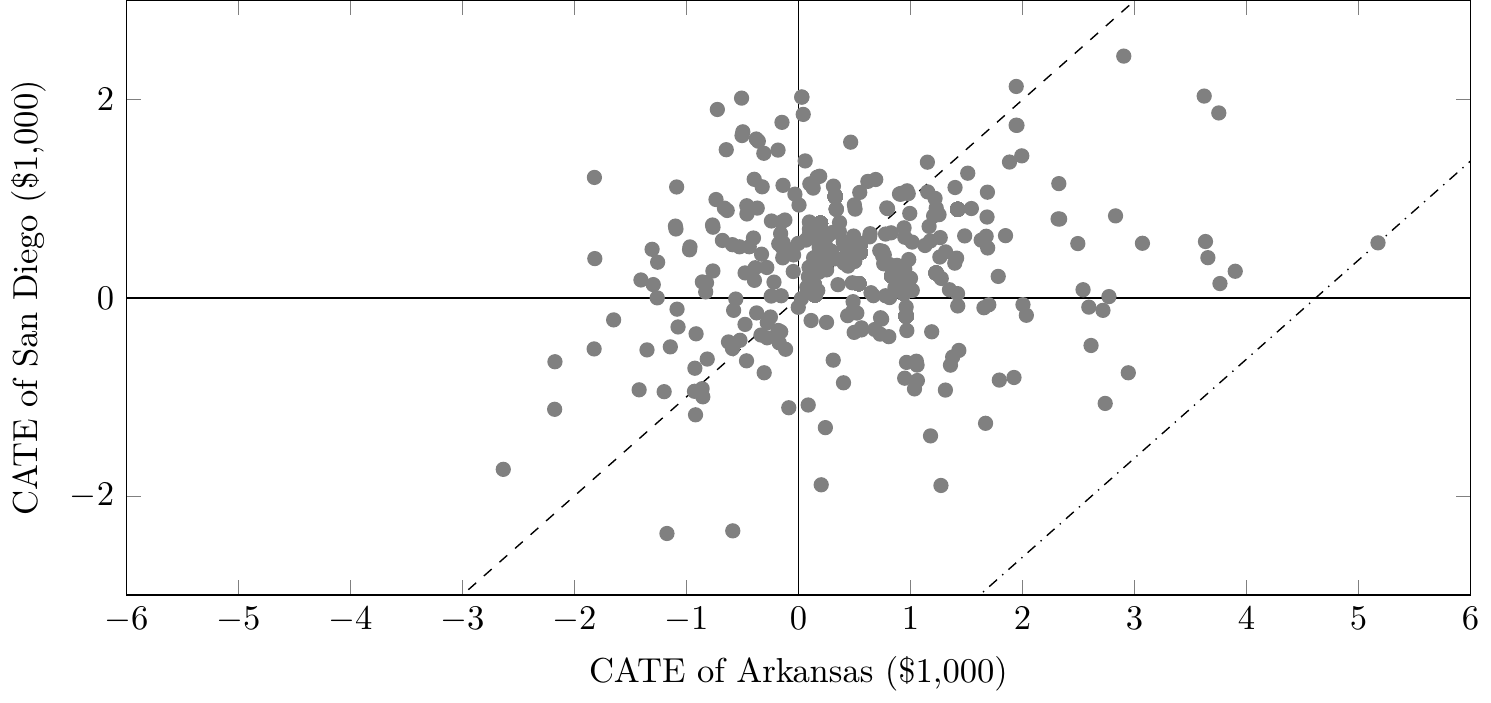}
    \begin{tablenotes}[flushleft]
    \footnotesize\item[] Notes: This figure illustrates the relation between Arkansas's conditional average treatment effect (CATE) function and San Diego's CATE function using a scatter plot. The CATE function of each location is estimated using the data from RCT conducted at the corresponding location. Each point represents an observed value of covariates in Arkansas's RCT, and its $x$-coordinate and $y$-coordinate are an evaluation of the estimated CATE function for Arkansas and San Diego at that value of covariates. The dashed line shows that the two CATEs are the same. Thus, the more distant the point is from this line, the more different the two CATEs are. The dotted-dashed line passes through the point farthest from the reference line. 
    \end{tablenotes}
    \end{threeparttable}
    \label{fig:difference-in-cate-functions-between-arkansas-and-sandiego}
\end{figure}
The source and target populations actually differ in terms of the covariate distribution and conditional distribution of potential outcomes. First, for the difference in the covariate distribution, \ifdefined\OVERALL\cref{tab:covariate-summary} \else\cref{appendix-tab:covariate-summary} in the supplementary materials \fi lists the covariates used in the following estimation and provides the summary statistics of covariates by RCTs. In addition, it compares the means of each covariate in the two RCTs. The result implies that the two populations differ in terms of means of the covariates. To show that that conditional distributions of potential outcomes are also different, I focus on the difference in the conditional average treatment effect (CATE) functions of the two populations. Specifically, I estimate the CATE function of each location using S-learner with the random forest as the base learner \citep[for a review of the CATE estimation, see][]{Kuenzel2019,Jacob2021}. Then, I evaluate the two estimated CATE functions at the observed values of covariates in Arkansas's RCT. Thus, the difference of the covariate distribution is essentially controlled, and any difference in the values of CATE derives from the difference in the conditional distribution of potential outcomes. \cref{fig:difference-in-cate-functions-between-arkansas-and-sandiego} visualizes the difference in the values of the CATE using a scatter plot. In the figure, each point represents an observation of covariates in Arkansas's RCT, and its $x$-coordinate and $y$-coordinate are an evaluation of the estimated CATE function for Arkansas and San Diego at that value of covariates. If a point is on the dashed line, the CATEs of the two locations are the same. In other words, the more distant the point is from this line, the more different the two CATEs are. From this figure, one can observe that the CATE functions of the two locations are substantively different. The maximum distance between the two CATEs are about \$4,618. In summary, the source and target populations are significantly different, and hence, the existing methods that assume two populations to be same are inappropriate.
\par
I apply the proposed approach to the current setting and estimate the DR-ITR. The outcome is the total earnings in two years after the start of the WORK or SWIM, and thus the outcome space is non-negative real numbers. The source CMR function is estimated by the random forest with tuned hyper parameters using R package \texttt{randomForest}. The ambiguity level $\delta$ in \cref{eq:wasserstein-ambiguity} varies in $0,1,\cdots,10$. Note that the case of $\delta = 0$ corresponds to the naive approach explained in \cref{sec:naive-approach}, which is included for comparison. The ITR class consists of decision trees of depth 2 constructed by the covariates except sex and race information. These two variables are excluded because this information should not be used owing to ethical concerns. The optimization of the tree is conducted using R package \texttt{policytree} \citep{Zhou2022}. After obtaining the ITRs, I evaluate their performance by estimating the total earnings that would be attained in two years if each ITR was implemented in Arkansas. Note that this corresponds to the target population's welfare. As the data from RCT conducted in Arkansas are actually available, it is possible to consistently estimate the welfare.
\par
\begin{figure}[t]
    \centering
    \caption{Performance of the distributionally robust individualized treatment rule for the target population}
    \begin{threeparttable}
    \includegraphics[]{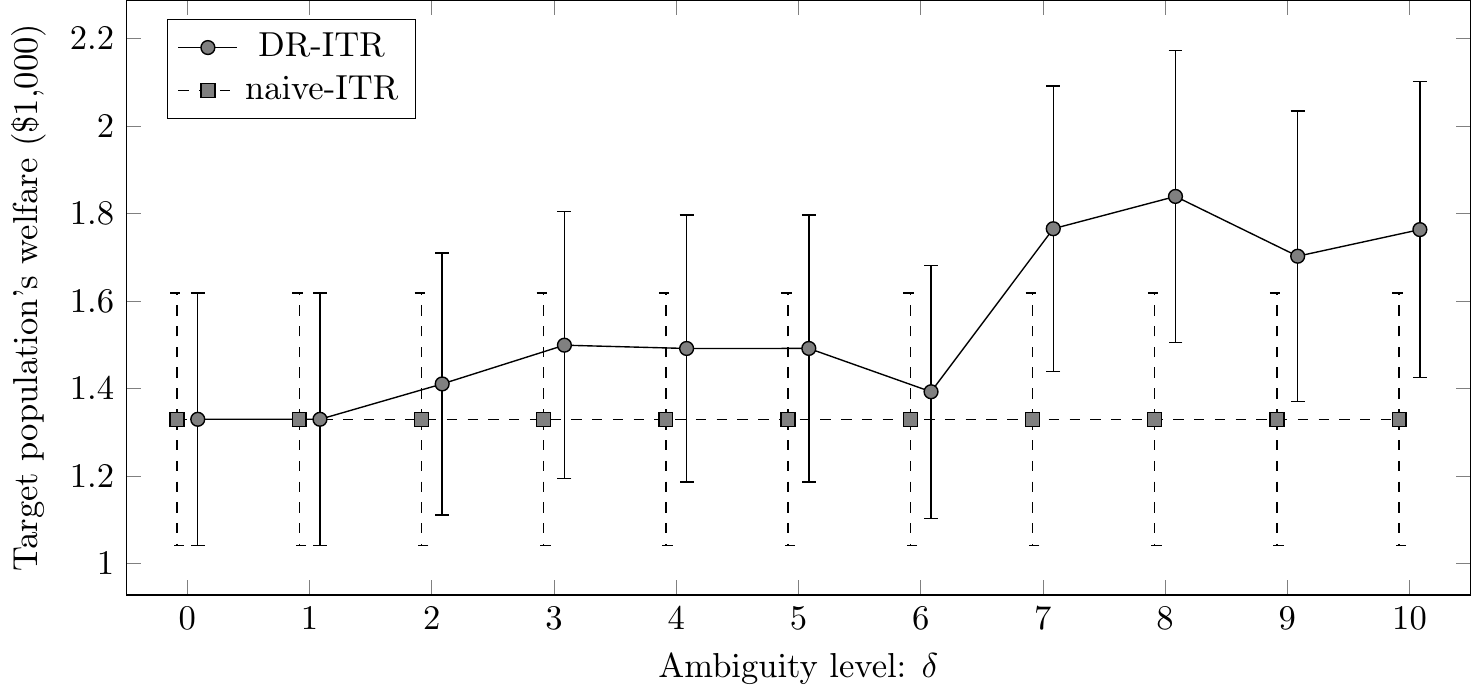}
    \begin{tablenotes}[flushleft]
    \footnotesize\item[] Notes: This figure summarizes the performance of the distributionally robust individualized treatment rule (DR-ITR) and naive-ITR in terms of the target population's welfare, which is the average earnings in two years that would be attained if an ITR was implemented in Arkansas. The solid line with a circle represents the DR-ITR, while the dashed line with a square represents the naive-ITR. For each line, each error bar indicates the 95\% confidence interval of the corresponding estimate. 
    \end{tablenotes}
    \end{threeparttable}
    \label{fig:performance-of-DR-ITR-at-the-target-population}
\end{figure}
\cref{fig:performance-of-DR-ITR-at-the-target-population} compares the performance of the DR-ITR with that of the naive-ITR. In the figure, the solid line with a circle represents the welfare of the DR-ITR, while the dashed line with a square represents that of the naive-ITR. Note that the naive-ITR does not depend on the value of $\delta$. It is clear that the DR-ITR outperforms the naive-ITR in terms of the point estimates of the target population's welfare by appropriately choosing the value of $\delta$. Specifically, the welfare due to the DR-ITR with positive ambiguity level $\delta$ is not less than that of the naive-ITR. Moreover, when $\delta = 9$, the t-test on the null hypothesis that the values of the target population's welfare due to the DR-ITR and naive-ITR are the same is rejected at the 5\% confidence level. 
\section{Conclusion}\label{sec:conclusion}
This study examined how one should determine a reasonable ITR when the population in which the estimated ITR is implemented differs from the population from which the available experimental or observational data are generated. Specifically, this study focuses on the unknown and unidentifiable difference in the conditional distribution of potential outcomes between the source and target populations, and proposes obtaining a DR-ITR that maximizes the worst-case value of welfare over a certain set of populations. In combination with the Wasserstein distance-based ambiguity set, the DR-ITR offers a simple intuition and estimation method. Additionally, it provides a justification for the naive approach, which assumes that the conditional distributions of potential outcomes are the same between the source and target populations. Thus, this study can be regarded as complementing the existing studies on this topic. In addition, an estimator for the distributionally robust policy is developed, and the theoretical analysis shows that the estimator has regret converging to zero.
\par
Several interesting questions are left unanswered. First, a policy maker often has access to multiple sets of experimental data, each of which is generated from a different population. It is obscure how the proposed DR-ITR can be extended to such multiple-source population scenarios. Second, in the population level, the current problem discussed in \cref{sec:model} can be understood as a two-person zero-sum game between a policymaker who chooses an ITR to maximize the target population's welfare and an adversary who chooses the target population to minimize the target population's welfare. However, this study does not consider whether a Nash equilibrium exists or what types of strategies constitute the equilibrium. I leave these questions for future work.

\ifdefined\OVERALL
\pagebreak

\maketitle

\appendix
\numberwithin{equation}{section}
\renewcommand{\theequation}{A.\arabic{equation}}
\renewcommand{\thetable}{A.\arabic{table}}

\paragraph{Structure of \ifSubfilesClassLoaded{this supplementary materials}{this appendix}}
\cref{sec:suppementary-information-on-empirical-application} provides the additional information on \ifSubfilesClassLoaded{\cref{main-sec:empirical-application}}{\cref{sec:empirical-application}}. \cref{sec:review-of-BM} gives the brief review of results of \citet{Blanchet2019a}. \cref{sec:proof-of-results} provides the proofs of results given in this study.
\paragraph{Additional notation}
Here, I introduce some notation that is exclusive to \ifSubfilesClassLoaded{this supplementary materials}{this appendix}. For an arbitrary space $\mathcal{S}$, I denote the identity mapping by $\mathrm{Id}$; that is, $\mathrm{Id}(s) = s$ for all $s \in \mathcal{S}$. In addition, for measurable spaces $(\mathcal{S},\mathscr{B}(\mathcal{S}),\mu)$ and $(\mathcal{T},\mathscr{B}(\mathcal{T}),\nu)$, the product measure over the corresponding product space is denoted by $\mu \otimes \nu$. 

\section{Supplementary Information on Empirical Application}\label{sec:suppementary-information-on-empirical-application}

\begin{table}[H]
\caption{Summary statistics of covariates}
\centering
\begin{threeparttable}
\begin{tabular*}{\textwidth}{l @{\extracolsep{\fill}} ccccccc}
\toprule
\multicolumn{1}{l}{Variable} & \multicolumn{3}{c}{Arkansas (WORK)} & \multicolumn{3}{c}{San Diego (SWIM)} & \\
\multicolumn{1}{l}{ } & \multicolumn{3}{c}{$N$ =  1,127 } & \multicolumn{3}{c}{$N$ =  3,211 } & \\
\cmidrule(l{3pt}r{3pt}){2-4}\cmidrule(l{3pt}r{3pt}){5-7}
& Min & Mean & Max & Min & Mean & Max & Diff. in means\\
&& [S.D.] &&& [S.D.] && (S.E.) \\
\midrule
\addlinespace
\multicolumn{8}{l}{Personal characteristics} \\
\addlinespace
\hspace{3mm} Female & $0.00$ & $0.98$ & $1.00$ & $0.00$ & $0.91$ & $1.00$ & $0.06$ \\
&& $[0.15]$ &&& $[0.28]$ && $(0.01)$ \\
\addlinespace
\hspace{3mm} High school dipl. & $0.00$ & $0.50$ & $1.00$ & $0.00$ & $0.56$ & $1.00$ & $-0.06$ \\
&& $[0.50]$ &&& $[0.50]$ && $(0.02)$ \\
\addlinespace
\hspace{3mm} Non-white & $0.00$ & $0.87$ & $1.00$ & $0.00$ & $0.68$ & $1.00$ & $0.19$ \\
&& $[0.34]$ &&& $[0.47]$ && $(0.01)$ \\
\addlinespace
\hspace{3mm} Never married & $0.00$ & $0.49$ & $1.00$ & $0.00$ & $0.30$ & $1.00$ & $0.19$ \\
&& $[0.50]$ &&& $[0.46]$ && $(0.02)$ \\
\addlinespace
\hspace{3mm} One child & $0.00$ & $0.39$ & $1.00$ & $0.00$ & $0.46$ & $1.00$ & $-0.06$ \\
&& $[0.49]$ &&& $[0.50]$ && $(0.02)$ \\
\addlinespace
\hspace{3mm} More than one child & $0.00$ & $0.58$ & $1.00$ & $0.00$ & $0.50$ & $1.00$ & $0.08$ \\
&& $[0.49]$ &&& $[0.50]$ && $(0.02)$ \\
\addlinespace
\addlinespace
\multicolumn{8}{l}{Employed prior to random assignment} \\
\addlinespace
\hspace{3mm} in quater 1 & $0.00$ & $0.15$ & $1.00$ & $0.00$ & $0.27$ & $1.00$ & $-0.11$ \\
&& $[0.36]$ &&& $[0.44]$ && $(0.01)$ \\
\addlinespace
\hspace{3mm} in quater 2 & $0.00$ & $0.16$ & $1.00$ & $0.00$ & $0.25$ & $1.00$ & $-0.09$ \\
&& $[0.37]$ &&& $[0.44]$ && $(0.01)$ \\
\addlinespace
\hspace{3mm} in quater 3 & $0.00$ & $0.14$ & $1.00$ & $0.00$ & $0.25$ & $1.00$ & $-0.11$ \\
&& $[0.35]$ &&& $[0.44]$ && $(0.01)$ \\
\addlinespace
\hspace{3mm} in quater 4 & $0.00$ & $0.15$ & $1.00$ & $0.00$ & $0.25$ & $1.00$ & $-0.10$ \\
&& $[0.36]$ &&& $[0.43]$ && $(0.01)$ \\
\addlinespace
\addlinespace
\multicolumn{8}{l}{Earnings (\$1,000) prior to random assignment} \\
\addlinespace
\hspace{3mm} in quater 1 & $0.00$ & $0.16$ & $4.96$ & $0.00$ & $0.39$ & $13.33$ & $-0.23$ \\
&& $[0.48]$ &&& $[1.00]$ && $(0.02)$ \\
\addlinespace
\hspace{3mm} in quater 2 & $0.00$ & $0.16$ & $3.75$ & $0.00$ & $0.40$ & $14.76$ & $-0.24$ \\
&& $[0.50]$ &&& $[1.02]$ && $(0.02)$ \\
\addlinespace
\hspace{3mm} in quater 3 & $0.00$ & $0.17$ & $4.42$ & $0.00$ & $0.39$ & $12.46$ & $-0.22$ \\
&& $[0.53]$ &&& $[1.02]$ && $(0.02)$ \\
\addlinespace
\hspace{3mm} in quater 4 & $0.00$ & $0.17$ & $3.57$ & $0.00$ & $0.38$ & $15.59$ & $-0.21$ \\
&& $[0.50]$ &&& $[1.03]$ && $(0.02)$ \\
\addlinespace
\bottomrule
\end{tabular*}
\label{tab:covariate-summary}
\begin{tablenotes}[flushleft]
\footnotesize\item Notes: This table presents the summary statistics of covariates that are used in the estimation of CMR functions and ITRs. Specifically, it shows minimum, mean, standard deviation (in bracket) and max of each covariate by RCTs. In addition, it also gives the difference in means and its standard error (in parenthesis). The covariates are the same as in \citet{Hotz2005} except that female dummy is included, and there are 14 covariates in total.
\end{tablenotes}
\end{threeparttable}
\end{table}

\section{Review of \texorpdfstring{\citet{Blanchet2019a}}{}}\label{sec:review-of-BM}

For completeness and for later use, I summarize the strong duality result due to \citet{Blanchet2019a} (hereafter I refer to their paper as BM for short).\par
Let $\mcs$ be a Polish space and let $\mu \in \msp(\mcs)$ be a reference probability measure. Consider functions $c:\mcs\times\mcs \to [0,+\infty]$ and $F:\mcs \to \mathbb{R}$ satisfying the following assumptions.
\begin{assumption}\label{assu:lower-semicontinuity-of-cost}
    The function $c:\mcs \times \mcs \to [0,+\infty]$ is a lower semi-continuous function satisfying $c(s,s) = 0$ for every $s \in \mcs$.
\end{assumption}
\begin{assumption}\label{assu:F-is-integrable}
    The function $F:\mcs \to \mathbb{R}$ is upper semi-continuous and integrable with respect to $\mu$.
\end{assumption}
The optimal transport cost associated with the cost function $c$ is defined as 
\begin{equation}
    C(\mu,\nu) := \inf_{\pi \in \Pi(\mu,\nu)} \int cd\pi
    \label{eq:def-optimal-transport-cost}
\end{equation}
for all $\mu,\nu\in\msp(\mcs)$. For fixed distribution $\mu \in \msp(\mcs)$ and $\delta \geq 0$, the quantity of interest is defined by
\begin{equation}
    \overline{I} 
    := \sup_{\nu \in \msp(\mcs)} \left\{\int Fd\nu \;\middle|\; C(\mu,\nu) \leq \delta\right\}.
    \label{eq:quantity-of-interest}
\end{equation}
It is well-known that under \cref{assu:lower-semicontinuity-of-cost} there exists the optimal transport plan $\pi \in \Pi(\mu,\nu)$ that attains the infimum in \cref{eq:def-optimal-transport-cost} \citep[see, e.g., Theorem~4.1 of][]{Villani2009}. In addition, the definition of transport plans implies that we have $\int Fd\nu = \int F(\tilde{s}) d\pi(s,\tilde{s})$ for every $\pi \in \Pi(\mu,\nu)$. Hence, \cref{eq:quantity-of-interest} can be rewritten as
\begin{equation}
    \overline{I} = \sup_{\pi \in \Phi_{\mu, \delta}} I(\pi)
    \label{eq:primal-problem}
\end{equation}
where
\begin{equation*}
    I(\pi) := \int F(\tilde{s})d\pi(s,\tilde{s})
    \text{ and } 
    \Phi_{\mu,\delta} := \left\{\pi \in \bigcup_{\nu \in \mathscr{P}(\mcs)} \Pi(\mu,\nu)
    \;\middle|\;
    \int cd\pi \leq \delta\right\}.
\end{equation*}
\Cref{eq:primal-problem} is referred as the primal problem and a transport plan that attains the supremum in primal problem, if it exists, are referred as a primal optimizer.\par
I next discuss the corresponding dual problem. Let me introduce some additional notations. For any Borel measure $\mu \in \msp(\mcs)$, let $\msb_{\mu}(\mcs)$ denote the completion of $\msb(\mcs)$ with respect to $\mu$. The universal $\sigma$-algebra is defined by $\msu(\mcs) = \cap_{\mu \in \msp(\mcs)} \msb_{\mu}(\mcs)$. In addition, let $\mathfrak{m}_{\mathscr{U}}(\mcs;\mathbb{R}\cup\{+\infty\})$ denote the collection of measurable function $\phi:(\mcs,\msu(\mcs)) \to (\mathbb{R}\cup\{+\infty\},\msb(\mathbb{R}\cup\{+\infty\}))$.\par
I move to the definition of dual problem. Let $C = \{(s,\tilde{s})\in\mcs\times\mcs \mid c(s,\tilde{s}) < \infty\}$. Note that under \cref{assu:lower-semicontinuity-of-cost} the set $C$ is Borel measurable. Define $\Lambda_{c,F}$ to be the collection of pairs $(\lambda, \phi)$ such that $\lambda$ is a non-negative real number, $\phi \in \mathfrak{m}_{\mathscr{U}}(\mcs;\mathbb{R}\cup\{+\infty\})$, and
\begin{equation*}
    \phi(s) + \lambda c(s,\tilde{s}) \geq F(\tilde{s}) \text{ for all }(s,\tilde{s}) \in C.
\end{equation*}
For every $(\lambda, \phi) \in \Lambda_{c,F}$, consider 
\begin{equation*}
    J(\lambda, \phi) := \lambda\delta + \int\phi\mu.
\end{equation*}
Then, the dual problem is defined by
\begin{equation}
    \underline{J} := \sup_{(\lambda, \phi) \in \Lambda_{c,F}} J(\lambda, \phi).
    \label{eq:dual-problem}
\end{equation}\par
As printed below, BM showed that the strong duality, $\overline{I} = \underline{J}$, holds under \cref{assu:lower-semicontinuity-of-cost,assu:F-is-integrable}. But, before stating the strong duality result, we observe that the weak duality, $\underline{J} \leq \overline{I}$, holds without the lower semi-continuity of $F$. Note that for any $\pi \in \Phi_{\mu,\delta}$ we have $\pi(C) = 1$ because $\int cd\pi$ is finite. Therefore, for any $\pi \in \Phi_{\mu,\delta}$ and $(\lambda, \phi) \in \Lambda_{c,F}$, it holds that
\begin{align*}
    J(\lambda, \phi) 
    &= \lambda \delta + \int_C \phi(s) d\pi(s,\tilde{s})\\
    &\geq \lambda \delta + \int_C \left(F(\tilde{s})-\lambda c(s,\tilde{s}) \right)d\pi(s,\tilde{s})\\
    &= \lambda \left(\delta - \int c(s,\tilde{s})d\pi(s,\tilde{s})\right) + \int F(\tilde{s})d\pi(s,\tilde{s})\\
    &\geq I(\pi),
\end{align*}
which implies $\underline{J} \geq \overline{I}$. In addition, we note that the dual problem can be further simplified as follows: For any $\lambda \geq 0$, define $\phi_\lambda: \mcs \to \mathbb{R}\cup\{+\infty\}$ such that
\begin{equation*}
    \phi_{\lambda}(s) := \sup_{\tilde{s}\in\mcs} \left\{F(\tilde{s}) - \lambda c(s,\tilde{s})\right\},
\end{equation*}
where $\lambda c(s,\tilde{s}) = +\infty$ when $\lambda = 0$ and $c(s,\tilde{s}) = +\infty$. The function $\phi_\lambda$ is actually an element of $\mathfrak{m}_{\mathscr{U}}(\mcs;\mathbb{R}\cup\{+\infty\})$ (see Section~4 of BM). Then, we have $J(\lambda,\phi_\lambda) \leq J(\lambda, \phi)$ for every $(\lambda, \phi) \in \Lambda_{c,F}$.\par
\begin{lemma}[{Theorem~1 of \citet{Blanchet2019a}}]\label{lem:blanchet-murthy-duality}
    Under \cref{assu:lower-semicontinuity-of-cost,assu:F-is-integrable}, 
    \begin{lemenum}[nosep]
        \item $\overline{I} = \underline{J}$ holds.
        \item There exists a dual optimizer of the form $(\lambda, \phi_\lambda)$ for some $\lambda \geq 0$. In addition, any feasible $\pi^* \in \Phi_{\mu,\delta}$ and $(\lambda^*,\phi_{\lambda^*}) \in \Lambda_{c,F}$ are primal and dual optimizers, satisfying $I(\pi^*) = J(\lambda^*, \phi_{\lambda^*})$, if and only if
        \begin{equation}
            \begin{aligned}
                &F(\tilde{s}) - \lambda^*c(s,\tilde{s}) = \sup_{s' \in \mcs} \left\{F(z') - \lambda^* c(s,s')\right\}, \text{ $\pi^*$ a.s., and }\\
                &\lambda^*\left(\int c(s,\tilde{s})d\pi(s,\tilde{s}) - \delta\right) = 0.
            \end{aligned}
            \label{eq:complementarity-constraints}
        \end{equation}
    \end{lemenum}
    Moreover, the primal optimizer $\pi^*$, if it exists, is unique if for $\mu$-almost every $s \in \mcs$, there is only one $\tilde{s}$ in $\mcs$ that attains the supremum in $\sup_{\tilde{s}\in\mcs} \left\{F(\tilde{s}) - \lambda^* c(s,\tilde{s})\right\}$.
\end{lemma}
The most significant implication of \cref{lem:blanchet-murthy-duality} is the worst-case expectation in \cref{eq:quantity-of-interest} can be calculated by solving the dual problem, i.e., 
\begin{equation*}
    \overline{I} = \inf_{\lambda \geq 0}\left\{\lambda \delta + E_{\mu}\left[\sup_{\tilde{s} \in \mcs}\left\{F(\tilde{s}) - \lambda c(S,\tilde{s})\right\}\right]\right\},
\end{equation*}
where $S$ is the random variable distributed according to $\mu$. Although the original problems in \cref{eq:quantity-of-interest,eq:primal-problem} are generally infinite dimensional problem, the dual problem is one dimensional and convex problem. Thus, the result makes the original problems tractable.\par 
\section{Proof of Results}\label{sec:proof-of-results}
\subsection{Proof of \texorpdfstring{\ifSubfilesClassLoaded{\cref{main-cor:characterize-ambiguity-set}}{\cref{cor:characterize-ambiguity-set}}}{}}
The corollary follows almost directly from Kantrovich-Rubinstein formula. Its proof can be found in \citet[Chapter~5]{Villani2009}.
\begin{lemma}[Kantrovich-Rubinstein formula]\label{lem:kantrovich-rubinstein-formula}
    Let $(\mcs,d_{\mcs})$ be a complete and separable metric space. For any $\mu,\nu \in \msp(\mcs)$, it holds 
    \begin{equation*}
        W_1(\mu,\nu) = \sup_{\lVert \psi\rVert_{\mathrm{lip}} \leq 1}\left\{\int_{\mcs} \psi d\mu - \int_{\mcs} \psi d\nu\right\},
    \end{equation*}
    where $\lVert \cdot \rVert_{\mathrm{lip}}$ denotes the lipschitz norm.
\end{lemma}
Given this result, I show \ifSubfilesClassLoaded{\cref{main-cor:characterize-ambiguity-set}}{\cref{cor:characterize-ambiguity-set}}.
\begin{proof}[Proof of \ifSubfilesClassLoaded{\cref{main-cor:characterize-ambiguity-set}}{\cref{cor:characterize-ambiguity-set}}]
    (i)(ii) They immediate follow from \cref{lem:kantrovich-rubinstein-formula} since the functions $(y_1,\cdots,y_d) \mapsto y_a$ or $(y_1,\cdots,y_d) \mapsto y_{a_1} - y_{a_2}$ are 1-lipschitz continuous.\\
    (iii)(iv) They immediate follow from combination of \ifSubfilesClassLoaded{\cref{main-corenum:charact-cmr,main-corenum:charact-cate}}{\cref{corenum:charact-cmr,corenum:charact-cate}} with the following inequalities:
    \begin{equation*}
        \begin{aligned}
            |\mathbf{E}_{\mbu}[Y_a] - \mathbf{E}_{\mbs}[Y_a]|
            &= |\mathbf{E}_{\mbt_X}[(m_{\mbu}(X,a)-m_{\mbs}(X,a)) + m_{\mbs}(X,a)]-\mathbf{E}_{\mbs_X}[m_{\mbs}(X,a)]|\\
            &\leq \mathbf{E}_{\mbt_X}[|m_{\mbu}(X,a)-m_{\mbs}(X,a)|] + |\mathbf{E}_{\mbt_X}[m_{\mbs}(X,a)]-\mathbf{E}_{\mbs_X}[m_{\mbs}(X,a)]|
        \end{aligned}
    \end{equation*}
    and 
    \begin{equation*}
        \begin{aligned}
            |\mathbf{E}_{\mbu}[Y_{a_1} - Y_{a_2}] - \mathbf{E}_{\mbs}[Y_{a_1} - Y_{a_2}]|
            &= |\mathbf{E}_{\mbt_X}[(m_{\mbu}(X,a_1)-m_{\mbu}(X,a_2)) - (m_{\mbs}(X,a_1)-m_{\mbs}(X,a_2)) \\
            &\quad + (m_{\mbs}(X,a_1)-m_{\mbs}(X,a_2))] - \mathbf{E}_{\mbs_X}[m_{\mbs}(X,a_1) - m_{\mbs}(X,a_2)]|\\
            &\leq \mathbf{E}_{\mbt_X}[|(m_{\mbu}(X,a_1)-m_{\mbu}(X,a_2)) - (m_{\mbs}(X,a_1)-m_{\mbs}(X,a_2))|]\\
            &\quad + |\mathbf{E}_{\mbt_X}[m_{\mbs}(X,a_1)-m_{\mbs}(X,a_2)] - \mathbf{E}_{\mbs_X}[m_{\mbs}(X,a_1) - m_{\mbs}(X,a_2)]|
        \end{aligned}
    \end{equation*}
\end{proof}
\subsection{Proof of \texorpdfstring{\ifSubfilesClassLoaded{\cref{main-thm:tractable-reformulation-dr-itr}}{\cref{thm:tractable-reformulation-dr-itr}}}{} and \texorpdfstring{\ifSubfilesClassLoaded{\cref{main-cor:existence-of-worst-case-population}}{\cref{cor:existence-of-worst-case-population}}}{}}
\begin{proof}[Proof of \ifSubfilesClassLoaded{\cref{main-thm:tractable-reformulation-dr-itr}}{\cref{thm:tractable-reformulation-dr-itr}}]
    To obtain the strong dual problem of \ifSubfilesClassLoaded{\cref{main-eq:worst-case-cond-welfare}}{\cref{eq:worst-case-cond-welfare}} using the result given in \cref{lem:blanchet-murthy-duality}, what one needs to do is to confirm that the assumptions in \cref{lem:blanchet-murthy-duality} are satisfied in the current setting. The satisfaction of \cref{assu:lower-semicontinuity-of-cost} is obvious. One can also easily verify \cref{assu:F-is-integrable} from \ifSubfilesClassLoaded{\cref{main-assuenum:source-outcome-almost-bounded}}{\cref{assuenum:source-outcome-almost-bounded}}. Then, the direct application of \cref{lem:blanchet-murthy-duality} implies that, it holds that
    \begin{equation}
        \underline{v}(x;g)
        =\max_{\lambda \geq 0} \left\{-\lambda\delta+\mathbf{E}_{{\mbs}_{(Y_{1},\cdots,Y_{d})\mid X=x}}\left[\inf_{(y_{1}',\cdots,y_{d}')\in\mathcal{Y}^{d}}\left( \sum_{a\in\mathcal{A}}y_{a}'\cdot1\{g(x)=a\}+\lambda\left|Y_{a}-y_{a}'\right|\right) \right]\right\}
        \label{eq:duality-by-BM2019}
    \end{equation}
    for each $x \in \supp(\mbt_X)$ and $g \in \mathcal{G}$.
    \par
    Next, I evaluate the local optimization problem in the expectation. For any $x\in\mathcal{X}$ and $\lambda\geq0$, it holds that 
    \begin{equation*}
        \inf_{(y_{1}',\cdots,y_{d}')\in\mathcal{Y}^{d}}\left\{ \sum_{a\in\mathcal{A}}y_{a}'\cdot1\{g(x)=a\}+\lambda|Y_{a}-y_{a}'|\right\} 
        =\sum_{a\in\mathcal{A}}\inf_{y_{a}'\in\mathcal{Y}}\left\{ y_{a}'\cdot1\{g(x)=a\}+\lambda|Y_{a}-y_{a}'|\right\}.
    \end{equation*}
    Hence, it suffices to solve the infimum by each $a\in\mathcal{A}$. Moreover, further calculation gives
    \begin{align*}
    \inf_{y_{a}'\in\mathcal{Y}}\left\{ y_{a}'\cdot1\{g(x)=a\}+\lambda|Y_{a}-y_{a}'|\right\}
    =\left(\{(1-\lambda)\inf\mathcal{Y}+\lambda Y_{a}\}\cdot1\{\lambda\in[0,1)\}+Y_{a}\cdot1\{\lambda\geq1\}\right)\cdot1\{g(x)=a\}.
    \end{align*}
    Then
    \begin{align*}
        &\mathbf{E}_{\mbs_{(Y_{1},\cdots,Y_{d})\mid X=x}}\left[\inf_{(y_{1}',\cdots,y_{d}')\in\mathcal{Y}^{d}}\left\{ \sum_{a\in\mathcal{A}}y_{a}'\cdot1\{g(x)=a\}+\lambda|Y_{a}-y_{a}'|\right\} \right]\\
        =&\left((1-\lambda)\inf\mathcal{Y}+\lambda\sum_{a\in\mathcal{A}}m_{\mbs}(x,a)\cdot1\{g(x)=a\}\right)\cdot1\{\lambda\in[0,1)\} +\left(\sum_{a\in\mathcal{A}}m_{\mbs}(x,a)\cdot1\{g(x)=a\}\right)\cdot1\{\lambda\geq1\}.
    \end{align*}
    By plugging the above equation in the dual problem \cref{eq:duality-by-BM2019} and solving the optimization problem with respect to $\lambda$, one obtains the closed form of $\underline{v}(x;g)$. The corresponding dual optimizer $\lambda^{*}$ is $1$ when $\sum_{a\in\mathcal{A}}m_{\mbs}(x,a)\cdot1\{g(x)=a\}-\delta\geq\inf\mathcal{Y}$, and $0$ otherwise.
\end{proof}
\begin{proof}[Proof of \ifSubfilesClassLoaded{\cref{main-cor:existence-of-worst-case-population}}{\cref{cor:existence-of-worst-case-population}}]
    By \cref{lem:blanchet-murthy-duality}, it suffices to find a pair of a coupling $\pi^*$ between $\mbs_{(Y_{1},\cdots,Y_{d})\mid X=x}$ and $t_{\#}\mbs_{(Y_{1},\cdots,Y_{d})\mid X=x}$ and a dual optimizer $\lambda^*\geq0$ with the following conditions satisfied: (a) the complementary slackness conditions given in \cref{lem:blanchet-murthy-duality} are satisfied; (b) it holds that $I(\pi^*)=\inf_{\Phi}I(\pi)$. Then the right-side marginal probability measure of $\pi^*$ is the worst-case conditional distribution.
    \par
    I first consider the case when the outcome space is convex and unbounded from below, and confirm that the pair $(\pi^*,\lambda^*)=\left(\mathrm{Id}_{\#}\mbs_{(Y_{1},\cdots,Y_{d})\mid X=x}\otimes t_{\#}\mbs_{(Y_{1},\cdots,Y_{d})\mid X=x},1\right)$ fulfills these requirements. One can verify condition (a) as follows: since
    \begin{equation*}
        \begin{aligned}
            \argmin_{(y_{1}',\cdots,y_{d}')\in\mathcal{Y}^{d}}\left\{ \sum_{a\in\mathcal{A}}y_{a}'\cdot1\{g(x)=a\}+\lambda^*\left|y_{a}-y_{a}'\right|\right\}
            & =\prod_{a\in\mathcal{A}}\argmin_{y_{a}'\in\mathcal{Y}}\left\{ y_{a}'\cdot1\{g(x)=a\}+\left|y_{a}-y_{a}'\right|\right\} \\
            & =\prod_{a\in\mathcal{A}}
            \begin{cases}
                (-\infty,y_{a}] & \text{if }g(x)=a\\
                \{y_{a}\} & \text{otherwise,}
            \end{cases}
        \end{aligned}
    \end{equation*}
    the coupling $\pi^*$ is concentrated on the set of local optimizers. In addition, the pair satisfies 
    \begin{equation*}
        \begin{aligned}
            \lambda^*\left(\mathbf{E}_{\pi^*}\left[\sum_{a\in\mathcal{A}}|Y_{a}-Y_{a}'|\right]-\delta\right)
            & =\mathbf{E}_{\pi^*}\left[\sum_{a\in\mathcal{A}}|Y_{a}-(Y_{a}\cdot1\{g(x)\neq a\}+(Y_{a}-\delta)\cdot1\{g(x)=a\})|\right]-\delta\\
            & =\sum_{a\in\mathcal{A}}\delta\cdot1\{g(x)=a\}-\delta=0.
        \end{aligned}
    \end{equation*}
    The satisfaction of condition (b) is obvious.
    \par
    I next prove the case when the outcome space is convex but bounded from below.
    Suppose that $m_{\mbs}(x,a)-\delta\geq\inf\mathcal{Y}$ and consider the pair $(\pi^*,\lambda^*)=\left(\mathrm{Id}_{\#}\mbs_{(Y_{1},\cdots,Y_{d})\mid X=x}\otimes t_{\#}\mbs_{(Y_{1},\cdots,Y_{d})\mid X=x},1\right)$.
    Note that one has $m_{\mbs}(x,a)-\inf\mathcal{Y}>0$ since $\delta>0$. One can verify condition (a) as follows: since the coupling $\pi^*$ is concentrated on the set 
    \begin{equation*}
        \argmin_{(y_{1}',\cdots,y_{d}')\in\mathcal{Y}^{d}}\left\{ \sum_{a\in\mathcal{A}}y_{a}'\cdot1\{g(x)=a\}+\lambda^*\left|y_{a}-y_{a}'\right|\right\}
        =\prod_{a\in\mathcal{A}}
        \begin{cases}
            [\inf\mathcal{Y},y_{a}] & \text{\text{if }}g(x)=a\\
            \{y_{a}\} & \text{otherwise}
        \end{cases}
    \end{equation*}
    and satisfies 
    \begin{equation*}
        \begin{aligned}
            &\lambda^*\left(\mathbf{E}_{\pi^*}\left[\sum_{a\in\mathcal{A}}|Y_{a}-Y_{a}'|\right]-\delta\right)\\
            =&\mathbf{E}_{\pi^*}\left[\sum_{a\in\mathcal{A}}\left|Y_{a}-\left(Y_{a}\cdot1\{g(x)\neq a\}+\left(Y_{a}-\frac{\delta(Y_{a}-\inf\mathcal{Y})}{m_{\mu}(x,a)-\inf\mathcal{Y}}\right)\cdot1\{g(x)=a\}\right)\right|\right]-\delta\\
            =&\mathbf{E}_{\pi^*}\left[\sum_{a\in\mathcal{A}}\frac{\delta(Y_{a}-\inf\mathcal{Y})}{m_{\mu}(x,a)-\inf\mathcal{Y}}\cdot1\{g(x)=a\}\right]-\delta\\
            =&0.
        \end{aligned}
    \end{equation*}
    The satisfaction of (b) is easy.
    \par
    Conversely, suppose that $m_{\mbs}(x,a)-\delta<\inf\mathcal{Y}$. Then, the pair $(\pi^*,\lambda^*)=(\mathrm{Id}_{\#}\mbs_{(Y_{1},\cdots,Y_{d})\mid X=x}\otimes t_{\#}\mbs_{(Y_{1},\cdots,Y_{d})\mid X=x},0)$ satisfies the conditions. One can verify condition (a) as follows: since the coupling $\pi^*$ is concentrated on the set
    \begin{align*}
        \argmin_{(y_{1}',\cdots,y_{d}')\in\mathcal{Y}^{d}}\left\{ \sum_{a\in\mathcal{A}}y_{a}'\cdot1\{g(x)=a\}+\lambda^*\left|y_{a}-y_{a}'\right|\right\}
        & =\prod_{a\in\mathcal{A}}\argmin_{y_{a}'\in[\inf\mathcal{Y},\infty)}\left\{ y_{a}'\cdot1\{g(x)=a\}\right\} \\
        & =\prod_{a\in\mathcal{A}}
        \begin{cases}
            \{\inf\mathcal{Y}\} & \text{if }g(x)=a\\{}
            [\inf\mathcal{Y},\infty) & \text{otherwise}
        \end{cases}
    \end{align*}
    The satisfaction of (b) is obvious.
\end{proof}

\subsection{Proof of \texorpdfstring{\ifSubfilesClassLoaded{\cref{main-thm:equiv-between-dr-naive}}{\cref{thm:equiv-between-dr-naive}}}{}}
The proof is divided into two cases, one in which \ifSubfilesClassLoaded{\cref{main-assuenum:source-fb-contained}}{\cref{assuenum:source-fb-contained}} holds and the other in which \ifSubfilesClassLoaded{\cref{main-assuenum:source-cmr-distant-from-bound}}{\cref{assuenum:source-cmr-distant-from-bound}} holds.\par
\textit{Case when \ifSubfilesClassLoaded{\cref{main-assuenum:source-fb-contained}}{\cref{assuenum:source-fb-contained}} holds}: Notice that $g_{\mbs}^{\text{FB}}$ satisfies
\begin{equation*}
    m_{\mbs}(x,g_{\mbs}^{\text{FB}}(x)) \geq m_{\mbs}(x,g(x))\quad \mbt_X\text{-a.s.}
\end{equation*}
for all $g \in \mcg$. As a result, it immediately follows that
\begin{equation*}
    g_{\mbs}^{\text{FB}} \in \argmax_{g \in \mcg} \mathbf{E}_{\mbt_X}\left[m_{\mbs}(X,g(X))\right].
\end{equation*}
Let $N :=\{x\in\mcx \mid g_{\text{naive-ITR}}^*(x) \notin \argmax_{a \in \mca} m_{\mbs}(x,a)\}$ and suppose that $\mbt_X(N) > 0$. Then, it follows that
\begin{align*}
    \mathbf{E}_{\mbt_X}[m_{\mbs}(X,g_{\mbs}^{\text{FB}}(X))]
    &= \mathbf{E}_{\mbt_X}[m_{\mbs}(X,g_{\mbs}^{\text{FB}}(X))\cdot 1\{X \in N\}] + \mathbf{E}_{\mbt_X}[m_{\mbs}(X,g_{\mbs}^{\text{FB}}(X))\cdot 1\{X \notin N\}]\\
    &> \mathbf{E}_{\mbt_X}[m_{\mbs}(X,g_{\text{naive-ITR}}^*(X))\cdot 1\{X \in N\}] + \mathbf{E}_{\mbt_X}[m_{\mbs}(X,g_{\text{naive-ITR}}^*(X))\cdot 1\{X \notin N\}]\\
    &= \mathbf{E}_{\mbt_X}[m_{\mbs}(X,g_{\text{naive-ITR}}^*(X))].
\end{align*}
This contradicts to the definition of the naive-ITR since $g_{\mbs}^{\text{FB}} \in \mcg$. Thus, one must have $\mbt_X(N) = 0$. This implies that
\begin{equation*}
    \max\{m_{\mbs}(x,g_{\text{naive-ITR}}^*(x))-\delta,\inf\mcy\} \geq \max\{m_{\mbs}(x,g(x))-\delta,\inf\mcy\}\quad \mbt_X\text{-a.s.}
\end{equation*}
for all $g \in \mcg$, which results in $g_{\text{naive-ITR}}^* \in \argmax_{g \in \mcg} \mathbf{E}_{\mbt_X}\left[\max\{m_{\mbs}(X,g(X))-\delta,\inf\mcy\}\right]$.
\par
\textit{Case when \ifSubfilesClassLoaded{\cref{main-assuenum:source-cmr-distant-from-bound}}{\cref{assuenum:source-cmr-distant-from-bound}}}: In this case,
\ifSubfilesClassLoaded{\cref{main-thm:tractable-reformulation-dr-itr}}{\cref{thm:tractable-reformulation-dr-itr}} implies that the worst-case conditional welfare of an ITR is always equal to $m_{\mbs}(x,g(x))-\delta$ for each $x\in\supp(\mbt_X)$ and $g \in \mcg$. Therefore, DR-ITR is the solution of the following optimization problem:
\begin{equation*}
    g_{\text{DR-ITR}}^{*}\in\argmax_{g\in\mathcal{G}}\mathbf{E}_{\mbt_{X}}\left[\sum_{a\in\mathcal{A}}m_{\mbs}(X,a)\cdot1\{g(X)=a\}\right]-\delta.
\end{equation*}
Because the second term of the objective function does not vary with ITR, any naive-ITR is also DR-ITR.
\subsection{Proof of \texorpdfstring{\ifSubfilesClassLoaded{\cref{main-thm:performance-guarantee}}{\cref{thm:performance-guarantee}}}{}}

First, I cite a version of Talagrand's inequality \citep{Talagrand1996} due to \citet{Bousquet2002,Bousquet2003}.
In \cref{lemma:talagrand-ineq-by-bousquet}, $h(x)=(1+x)\log(1+x)-x$.
\begin{lemma}\label{lemma:talagrand-ineq-by-bousquet}
    Assume that $X_{i},i=1,\cdots,n$ are independent random variables.
    Let $\mathcal{F}$ be a countable set of functions from $\mathcal{X}$ to $\mathbb{R}$ and assume that all functions $f$ in $\mathcal{F}$ are measurable, square integrable and satisfy $\mathbf{E}[f(X_{i})]=0$.
    Assume $\sup_{f\in\mathcal{F}}\esssup|f|\leq U$ and denote 
    \begin{equation*}
        Z=\sup_{f\in\mathcal{F}}\left|\sum_{i=1}^{n}f(X_{i})\right|.
    \end{equation*}
    Let $\sigma$ be a positive real number such that $n\sigma^{2}\geq\sum_{i=1}^{n}\sup_{f\in\mathcal{F}}\mathbf{E}[f^{2}(X_{i})]$, then for all $x\geq0$, we have 
    \begin{equation*}
        \mathbf{P}(Z\geq\mathbf{E}[Z]+x)\leq\exp\left(-\frac{v}{U}h\left(\frac{xU}{v}\right)\right)
    \end{equation*}
    with $v=n\sigma^{2}+2U\mathbf{E}[Z]$ and also 
    \begin{equation*}
        \mathbf{P}\left(Z\geq\mathbf{E}[Z]+\sqrt{2xv}+\frac{xU}{3}\right)\leq\exp(-x).
    \end{equation*}
\end{lemma}
For ease of notation, let $\phi(x,a) := (m_{\mbs}(x,a)-\delta)\vee\inf\mathcal{Y}$ and $\widehat{\phi}(x,a) := (\widehat{m}_{\mbs}(x,a)-\delta)\vee\inf\mathcal{Y}$.
In addition, define $\bm{g}(x) := (g(x,1),\cdots,g(x,d))^\top$, $\bm{\phi}(x) := (\phi(x,1),\cdots,\phi(x,d))^{\top}$, and $\widehat{\bm{\phi}}(x) := (\widehat{\phi}(x,1),\cdots,\widehat{\phi}(x,d))^{\top}$.
Notice that with these notations it holds that $\underline{V}(g)=\mathbf{E}_{\nu_{X}}[\langle \bm{g}(X),\bm{\phi}(X)\rangle]$ and $\widehat{\underline{V}}(g) = \frac{1}{n_t}\sum_{j=1}^{n_t}\langle\bm{g}(X_j^t),\widehat{\bm{\phi}}(X_j^t)\rangle$, where $\langle\cdot,\cdot\rangle$ represents the usual inner product.
\par
Define 
\begin{equation*}
    \widetilde{\underline{V}}(g)=\frac{1}{n_{t}}\sum_{j=1}^{n_{t}}\langle \bm{g}(X_{j}^{t}),\bm{\phi}(X_{j}^{t})\rangle.
\end{equation*}
Following the proof technique of \citet{Zhou2022}, our proof strategy is based on the following inequality: 
\begin{align}
    & R_{\mathrm{DRPL}}(\widehat{g}_{\mathrm{DRP}})\nonumber \\
    =&\widehat{\underline{V}}(g_{\mathrm{DRPL}}^{*})-\widehat{\underline{V}}(\widehat{g}_{\mathrm{DRPL}})+\underline{V}(g_{\mathrm{DRPL}}^{*})-\underline{V}(\widehat{g}_{\mathrm{DRPL}})-\left(\widehat{\underline{V}}(g_{\mathrm{DRPL}}^{*})-\widehat{\underline{V}}(\widehat{g}_{\mathrm{DRPL}})\right)\nonumber \\
    \leq&\underline{V}(g_{\mathrm{DRPL}}^{*})-\underline{V}(\widehat{g}_{\mathrm{DRPL}})-\left(\widehat{\underline{V}}(g_{\mathrm{DRPL}}^{*})-\widehat{\underline{V}}(\widehat{g}_{\mathrm{DRPL}})\right)\nonumber \\
    \leq&\sup_{g,g'\in\mathcal{G}}\left|\left(\underline{V}(g)-\underline{V}(g')\right)-\left(\widehat{\underline{V}}(g)-\widehat{\underline{V}}(g')\right)\right|\nonumber \\
    \leq&\sup_{g,g'\in\mathcal{G}}\left|\left(\underline{V}(g)-\underline{V}(g')\right)-\left(\widetilde{\underline{V}}(g)-\widetilde{\underline{V}}(g')\right)\right|+\sup_{g,g'\in\mathcal{G}}\left|\left(\widetilde{\underline{V}}(g)-\widetilde{\underline{V}}(g')\right)-\left(\widehat{\underline{V}}(g)-\widehat{\underline{V}}(g')\right)\right|,\label{eq:decomposition-of-regret}
\end{align}
where the first inequality follows by definition of $\widehat{g}_{\text{DR-ITR}}$ and the third inequality follows by triangle inequality.
The first term in \cref{eq:decomposition-of-regret} represents the approximation error generated by using the empirical distribution instead of the true target covariate distribution.
On the other hand, the second term in \cref{eq:decomposition-of-regret} represents the estimation error generated by using the estimate $\widehat{m}_{\mbs}(x,a)$ instead of $m_{\mbs}(x,a)$.
In what follows, we bound the first and second terms separately. 
\par
As the above inequality \cref{eq:decomposition-of-regret} indicates, we focus on a specific function class constructed from the difference of arbitrary two policies. More specifically, we focus on a class of functions, $\mathcal{G}^{D}=\{\langle \bm{g}(\cdot)-\bm{g}'(\cdot),\cdot\rangle\mid g,g'\in\mathcal{G}\}$. Each element of $\mathcal{G}^{D}$ is a function that takes $X\in\mathcal{X}$ and a $d$-dimensional vector $\Gamma\in\mathbb{R}^{d}$ as input and outputs $\langle \bm{g}(X)-\bm{g}'(X),\Gamma\rangle$. We define the empirical Rademacher complexity of $\mathcal{G}^{D}$ as 
\begin{equation*}
\mathcal{R}_{n_{t}}(\mathcal{G}^{D};\{X_{j}^{t},\Gamma_{j}\}_{j=1}^{n_{t}})=\mathbf{E}\left[\sup_{g,g'\in\mathcal{G}}\frac{1}{n_{t}}\left|\sum_{j=1}^{n_{t}}\sigma_{j}\langle \bm{g}(X_{j}^{t})-\bm{g}'(X_{j}^{t}),\Gamma_{j}\rangle\right|\middle|\{X_{j}^{t},\Gamma_{j}\}_{j=1}^{n_{t}}\right],
\end{equation*}
where $\sigma_{j}$ is the Rademacher random variable such that it is independent from any other random variables. Then the Rademacher complexity of $\mathcal{G}^{D}$ is defined as $\mathcal{R}_{n_{t}}(\mathcal{G}^{D})=\mathbf{E}[\mathcal{R}_{n_{t}}(\mathcal{G}^{D};\{X_{j}^{t},\Gamma_{j}\}_{j=1}^{n_{t}})]$. We borrow the following result about the bound on the Rademacher complexity of $\mathcal{G}^{D}$ from \citet{Zhou2022}. 
\begin{lemma}
\label{lemma:bound-on-rademacher-complexity}Let $\{\Gamma_{j}\}_{j=1}^{n_{t}}$ be i.i.d. random vectors with bounded support. Then under \ifSubfilesClassLoaded{\cref{main-assu:policy-class}}{\cref{assu:policy-class}}, it holds that 
\begin{equation*}
\mathcal{R}_{n_{t}}(\mathcal{G}^{D})=27.2\sqrt{2}(\kappa(\mathcal{G})+8)\sqrt{\frac{V_{*}}{n_{t}}}+o\left(\frac{1}{\sqrt{n_{t}}}\right),
\end{equation*}
where $V_{*}=\sup_{g,g'\in\mathcal{G}}\mathbf{E}[\langle \bm{g}(X_{1}^{t})-\bm{g}'(X_{1}^{t}),\Gamma_{1}\rangle^{2}]$. 
\end{lemma}
\begin{lemma}
\label{lemma:general-high-probability-bound}Let $\{\Gamma_{j}\}_{j=1}^{n_{t}}$ be i.i.d. random vectors with bounded support and suppose that \ifSubfilesClassLoaded{\cref{main-assu:policy-class}}{\cref{assu:policy-class}} is in force. Then, for any $\epsilon\in(0,1)$, with probability at least $1-\epsilon$, it holds that 
\begin{align*}
&\sup_{g,g'\in\mathcal{G}}\left|\frac{1}{n_{t}}\sum_{j=1}^{n_{t}}\left(\langle \bm{g}(X_{j}^{t})-\bm{g}'(X_{j}^{t}),\Gamma_{j}\rangle-\mathbf{E}\left[\langle \bm{g}(X_{j}^{t})-\bm{g}'(X_{j}^{t}),\Gamma_{j}\rangle\right]\right)\right|\\
\leq&\left(54.4\sqrt{2}(\kappa(\mathcal{G})+8)+\sqrt{2\log\frac{1}{\epsilon}}\right)\sqrt{\frac{\sup_{g,g'\in\mathcal{G}}\mathbf{E}\left[\langle \bm{g}(X_{1}^{t})-\bm{g}'(X_{1}^{t}),\Gamma_{1}\rangle^{2}\right]}{n_{t}}}+o\left(\frac{1}{\sqrt{n_{t}}}\right).
\end{align*}
\end{lemma}
\begin{proof}[Proof of \cref{lemma:general-high-probability-bound}]
In what follows, we denote by $f_{g,g'}(X_{j}^{t},\Gamma_{j})$ the summand in the left-hand side of the above inequality.
First, we derive a bound on $\mathbf{E}[Z]/n_{t}$ where
\begin{equation*}
    Z=\sup_{g,g'\in\mathcal{G}}|\sum_{j=1}^{n_{t}}f_{g,g'}(X_{j}^{t},\Gamma_{j})|.
\end{equation*}
By the usual symmetrization argument \citep[see, e.g., Proposition 4.11 of][]{Wainwright2019} and applying \cref{lemma:bound-on-rademacher-complexity}, we obtain 
\begin{align*}
    \frac{1}{n_{t}}\mathbf{E}\left[Z\right] & \leq2\mathbf{E}\left[\mathbf{E}\left[\sup_{g,g'\in\mathcal{G}}\frac{1}{n_{t}}\left|\sum_{j=1}^{n_{t}}\sigma_{j}\langle \bm{g}(X_{j}^{t})-\bm{g}'(X_{j}^{t}),\Gamma_{j}\rangle\right|\middle|\{X_{j}^{t},\Gamma_{j}\}_{j=1}^{n_{t}}\right]\right]\\
    & =54.4\sqrt{2}(\kappa(\mathcal{G})+8)\sqrt{\frac{\sup_{g,g'\in\mathcal{G}}\mathbf{E}[\langle \bm{g}(X_{1}^{t})-\bm{g}'(X_{1}^{t}),\Gamma_{1}\rangle^{2}]}{n_{t}}}+o\left(\frac{1}{\sqrt{n_{t}}}\right).
\end{align*}
Importantly, this inequality implies that $\mathbf{E}[Z]/n_{t}=O(n_{t}^{-1/2}).$
\par
Next, we apply \cref{lemma:talagrand-ineq-by-bousquet} to obtain a high probability bound on $Z$.
We begin with defining relevant objects and confirming whether the conditions are satisfied.
Since $\Gamma_{j}$ is bounded almost surely, there exists $\widetilde{U}>0$ such that $\lVert\Gamma_{j}\rVert_{2}^{2}\leq\widetilde{U}$ a.s..
Then Cauchy-Schwarz inequality implies that $|f_{g,g'}(X_{j}^{t},\Gamma_{j})|\leq U$ a.s., where $U=2\sqrt{2}\widetilde{U}$.
In addition, if $\sigma^{2}=\sup_{g,g'\in\mathcal{G}}\mathbf{E}[\langle \bm{g}(X_{1}^{t})-\bm{g}'(X_{1}^{t}),\Gamma_{1}\rangle^{2}]$, $\sigma^{2}$ satisfies the condition since $(X_{j}^{t},\Gamma_{j}),j=1,\cdots,n_{t}$ are i.i.d..
Hence, \cref{lemma:talagrand-ineq-by-bousquet} implies that with probability at least $1-\epsilon$, we have 
\begin{equation*}
    \frac{1}{n_{t}}Z\leq\frac{1}{n_{t}}\mathbf{E}[Z]+\frac{1}{n_{t}}\sqrt{2\left(n_{t}\sigma^{2}+2U\mathbf{E}[Z]\right)\log\frac{1}{\epsilon}}+\frac{U}{3n_{t}}\log\frac{1}{\epsilon}.
\end{equation*}
Combining the bound on $\mathbf{E}[Z]/n_{t}$ with the above inequality, we finish the proof as follows:
\begin{align*}
    \frac{1}{n_{t}}Z\leq & \frac{1}{n_{t}}\mathbf{E}[Z]+\sqrt{2\left(\log\frac{1}{\epsilon}\right)\frac{\sigma^{2}}{n_{t}}}+\sqrt{\frac{4U}{n_{t}}\log\frac{1}{\epsilon}\frac{\mathbf{E}[Z]}{n_{t}}}+O\left(\frac{1}{n_{t}}\right)\\
    = & \frac{1}{n_{t}}\mathbf{E}[Z]+\sqrt{2\log\frac{1}{\epsilon}}\sqrt{\frac{\sigma^{2}}{n_{t}}}+\sqrt{\frac{4U}{n_{t}}\log\frac{1}{\epsilon}O\left(\frac{1}{\sqrt{n_{t}}}\right)}+O\left(\frac{1}{n_{t}}\right)\\
    = & \frac{1}{n_{t}}\mathbf{E}[Z]+\sqrt{2\log\frac{1}{\epsilon}}\sqrt{\frac{\sigma^{2}}{n_{t}}}+o\left(\frac{1}{\sqrt{n_{t}}}\right).
    \end{align*}
\end{proof}
This lemma immediately gives us an upper bound on the first term in \cref{eq:decomposition-of-regret}.
\begin{corollary}\label{corollary:convergence-rate-of-approximation-error}
    Suppose that \ifSubfilesClassLoaded{\cref{main-assu:relation-source-target,main-assu:policy-class}}{\cref{assu:relation-source-target,assu:policy-class}} are in force. Then, for any $\epsilon\in(0,1)$, with probability at least $1-\epsilon$, it holds that 
    \begin{align*}
        \sup_{g,g'\in\mathcal{G}}\left|\left(\underline{V}(g)-\underline{V}(g')\right)-\left(\widetilde{\underline{V}}(g)-\widetilde{\underline{V}}(g')\right)\right|
        \leq\left(54.4\sqrt{2}(\kappa(\mathcal{G})+8)+\sqrt{2\log\frac{1}{\epsilon}}\right)\sqrt{\frac{2d(M+\delta)^{2}}{n_{t}}}+o\left(\frac{1}{\sqrt{n_{t}}}\right),
    \end{align*}
    where $M<\infty$ is an almost sure bound of $m_{\mbs}(x,a)$.
\end{corollary}
\begin{proof}[Proof of \cref{corollary:convergence-rate-of-approximation-error}]
It suffices to apply \cref{lemma:general-high-probability-bound} with identification of $\Gamma_{j}=\bm{\phi}(X_{j}^{t})$.
\end{proof}
For the second term in \cref{eq:decomposition-of-regret}, we have the following lemma.
\begin{lemma}\label{lemma:convergence-rate-of-estimation-error}
    Suppose that \ifSubfilesClassLoaded{\cref{main-assu:relation-source-target,main-assu:dgp,main-assu:policy-class}}{\cref{assu:relation-source-target,assu:dgp,assu:policy-class}} are in force. Then, for any $\epsilon\in(0,1)$, with probability at least $1-\epsilon$, it holds that
    \begin{equation*}
    \sup_{g,g'\in\mathcal{G}}\left|\left(\widetilde{\underline{V}}(g)-\widetilde{\underline{V}}(g')\right)-\text{\ensuremath{\left(\widehat{\underline{V}}(g)-\widehat{\underline{V}}(g')\right)}}\right|=\sqrt{\frac{4C_{1}}{\epsilon\psi_{n_{s}}}}+o\left(\frac{1}{\sqrt{n_{t}}}\right).
    \end{equation*}
\end{lemma}
\begin{proof}[Proof of \cref{lemma:convergence-rate-of-estimation-error}]
From triangle inequality, we have 
\begin{align}
    \left|\left(\widetilde{\underline{V}}(g)-\widetilde{\underline{V}}(g')\right)-\text{\ensuremath{\left(\widehat{\underline{V}}(g)-\widehat{\underline{V}}(g')\right)}}\right|
    =&\left|\frac{1}{n_{t}}\sum_{j=1}^{n_{t}}\left\langle \bm{g}(X_{j}^{t})-\bm{g}'(X_{j}^{t}),\widehat{\bm{\phi}}(X_{j}^{t})-\bm{\phi}(X_{j}^{t})\right\rangle \right|\\
    \leq&\frac{1}{n_{t}}\left|\sum_{j=1}^{n_{t}}\left(\widehat{f}_{g,g'}(X_{j}^{t})-\mathbf{E}\left[\widehat{f}_{g,g'}(X_{j}^{t})\middle|\widehat{\phi}\right]\right)\right|+\frac{1}{n_{t}}\left|\sum_{j=1}^{n_{t}}\mathbf{E}\left[\widehat{f}_{g,g'}(X_{j}^{t})\middle|\widehat{\phi}\right]\right|\label{eq:decomposition-of-estimation-error}
\end{align}
where $\widehat{f}_{g,g'}(X_{j}^{t})=\langle \bm{g}(X_{j}^{t})-\bm{g}'(X_{j}^{t}),\widehat{\bm{\phi}}(X_{j}^{t})-\bm{\phi}(X_{j}^{t})\rangle$. We first give an upper bound on the first term of \cref{eq:decomposition-of-estimation-error}.
Since $\widehat{\phi}(\cdot)$ is computed using only the data $\{(Y_{i},A_{i},X_{i}^{s})\}_{i=1}^{n_{s}}$, \ifSubfilesClassLoaded{\cref{main-assu-item:independence}}{\cref{assu-item:independence}} implies that $\{\widehat{f}_{g,g'}(X_{j}^{t})\}_{j=1}^{n_{t}}$ are i.i.d. random variables conditional on $\widehat{\phi}$.
In addition, $\widehat{\bm{\phi}}(X_{j}^{t})-\bm{\phi}(X_{j}^{t})$ is almost surely bounded from \ifSubfilesClassLoaded{\cref{main-assu-item:convergence-rate-of-mse}}{\cref{assu-item:convergence-rate-of-mse}}.
Hence, if we write $Z_{1}=\sup_{g,g'\in\mathcal{G}}\frac{1}{n_{t}}|\sum_{j=1}^{n_{t}}\widehat{f}_{g,g'}(X_{j}^{t})-\mathbf{E}[\widehat{f}_{g,g'}(X_{j}^{t})\mid\widehat{\phi}]|$, \cref{lemma:general-high-probability-bound} with identification of $\Gamma_{j}=\widehat{\bm{\phi}}(X_{j}^{t})-\bm{\phi}(X_{j}^{t})$ implies the probability that inequality, 
\begin{equation}
    Z_{1}\leq\left(54.4\sqrt{2}(\kappa(\mathcal{G})+8)+\sqrt{2\log\frac{2}{\epsilon}}\right)\sqrt{\frac{\sup_{g,g'\in\mathcal{G}}\mathbf{E}[\widehat{f}_{g,g'}^{2}(X_{1}^{t})\mid\widehat{\phi}]}{n_{t}}}+o\left(\frac{1}{\sqrt{n_{t}}}\right),\label{eq:deviation-from-mean-conditional-on-phi}
\end{equation}
does not hold is less than $\epsilon/2$. Moreover, applying Cauchy-Schwarz inequality to $\widehat{f}_{g,g'}^{2}(X_{1}^{t})$ implies 
\begin{equation*}
    Z_{1}\leq\left(54.4\sqrt{2}(\kappa(\mathcal{G})+8)+\sqrt{2\log\frac{2}{\epsilon}}\right)\sqrt{\frac{2\mathbf{E}[\lVert\widehat{\bm{\phi}}(X_{1}^{t})-\bm{\phi}(X_{1}^{t})\rVert_{2}^{2}\mid\widehat{\phi}]}{n_{t}}}+o\left(\frac{1}{\sqrt{n_{t}}}\right).
\end{equation*}
\par
We next bound the second term of \cref{eq:decomposition-of-estimation-error}. Since $\{\widehat{f}_{g,g'}(X_{j}^{t})\}_{j=1}^{n_{t}}$ are i.i.d. random variables given $\widehat{\phi}$, if we write $Z_{2}=\sup_{g,g'\in\mathcal{G}}\frac{1}{n_{t}}|\sum_{j=1}^{n_{t}}\mathbf{E}[\widehat{f}_{g,g'}(X_{j}^{t})\mid\widehat{\phi}]|$, we have $Z_{2}=\sup_{g,g'\in\mathcal{G}}|\mathbf{E}[\widehat{f}_{g,g'}(X_{1}^{t})\mid\widehat{\phi}]|$. Moreover, we have the following series of inequalities: 
\begin{align*}
Z_{2} & \leq\sup_{g,g'\in\mathcal{G}}\mathbf{E}\left[\left|\widehat{f}_{g,g'}(X_{1}^{t})\right|\;\middle|\;\widehat{\phi}\right]\\
 & \leq\mathbf{E}\left[\left(2\lVert\widehat{\bm{\phi}}(X_{1}^{t})-\bm{\phi}(X_{1}^{t})\rVert_{2}^{2}\right)^{1/2}\;\middle|\;\widehat{\phi}\right]\\
 & \leq\sqrt{2\mathbf{E}\left[\lVert\widehat{\bm{\phi}}(X_{1}^{t})-\bm{\phi}(X_{1}^{t})\rVert_{2}^{2}\;\middle|\;\widehat{\phi}\right]},
\end{align*}
where the first and third inequality follow from Jensen's inequality, and the second inequality follows from Cauchy-Schwarz inequalilty. By combining this inequality with \cref{eq:deviation-from-mean-conditional-on-phi} we obtain
\begin{align*}
Z_{1}+Z_{2} & \leq\left(54.4\sqrt{2}(\kappa(\mathcal{G})+8)+\sqrt{2\log\frac{2}{\epsilon}}\right)\sqrt{\frac{2}{n_{t}}}\sqrt{\mathbf{E}\left[\lVert\widehat{\bm{\phi}}(X_{1}^{t})-\bm{\phi}(X_{1}^{t})\rVert_{2}^{2}\;\middle|\;\widehat{\phi}\right]}\\
 & \qquad+\sqrt{2\mathbf{E}\left[\lVert\widehat{\bm{\phi}}(X_{1}^{t})-\bm{\phi}(X_{1}^{t})\rVert_{2}^{2}\;\middle|\;\widehat{\phi}\right]}+o\left(\frac{1}{\sqrt{n_{t}}}\right).\\
\end{align*}
Since $\lVert\widehat{\bm{\phi}}(X_{1}^{t})-\bm{\phi}(X_{1}^{t})\rVert_{2}^{2}\leq\sum_{a\in\mathcal{A}}|\widehat{m}(X_{1}^{t},a)-m_{\mbs}(X_{1}^{t},a)|^{2}$, Markov inequality and \ifSubfilesClassLoaded{\cref{main-assu-item:convergence-rate-of-mse}}{\cref{assu-item:convergence-rate-of-mse}} imply that 
\begin{equation*}
\mathbf{P}\left(\mathbf{E}\left[\lVert\widehat{\bm{\phi}}(X_{1}^{t})-\bm{\phi}(X_{1}^{t})\rVert_{2}^{2}\;\middle|\;\widehat{\phi}\right]>\frac{2C_{1}}{\epsilon\psi_{n_{s}}}\right)<\frac{\epsilon}{2}.
\end{equation*}
for sufficiently large $n_{s}$. Combining this high probability bound on $\mathbf{E}[\lVert\widehat{\bm{\phi}}(X_{1}^{t})-\bm{\phi}(X_{1}^{t})\rVert_{2}^{2}\mid\widehat{\phi}]$ with the bound on $Z_{1}+Z_{2}$ implies that with probability at least $1-\epsilon$ we have
\begin{align*}
Z_{1}+Z_{2} & \leq\left(54.4\sqrt{2}(\kappa(\mathcal{G})+8)+\sqrt{2\log\frac{2}{\epsilon}}\right)\sqrt{\frac{4C_{1}}{\epsilon n_{t}\psi_{n_{s}}}}+\sqrt{\frac{4C_{1}}{\epsilon\psi_{n_{s}}}}+o\left(\frac{1}{\sqrt{n_{t}}}\right),\\
 & =\sqrt{\frac{4C_{1}}{\epsilon\psi_{n_{s}}}}+o\left(\frac{1}{\sqrt{n_{t}}}\right),
\end{align*}
which concludes the proof.
\end{proof}

\begin{proof}[Proof of \ifSubfilesClassLoaded{\cref{main-thm:performance-guarantee}}{\cref{thm:performance-guarantee}}]
    By applying \cref{corollary:convergence-rate-of-approximation-error} and \cref{lemma:convergence-rate-of-estimation-error} to inequality \cref{eq:decomposition-of-regret}, it holds that
    \begin{equation*}
        R_{\mathrm{\text{DRPL}}}(\widehat{g}_{\mathrm{\text{DRPL}}})\leq\left(54.4\sqrt{2}(\kappa(\mathcal{G})+8)+\sqrt{2\log\frac{2}{\epsilon}}\right)\sqrt{\frac{2d(M+\delta)^{2}}{n_{t}}}+\sqrt{\frac{8C_{1}}{\epsilon\psi_{n_{s}}}}+o\left(\frac{1}{\sqrt{n_{t}}}\right)
    \end{equation*}
    with probability at least $1-\epsilon$. 
\end{proof}

\ifSubfilesClassLoaded{%
\bibliography{reference}%
}{}

\end{document}

\fi

\bibliography{reference}

\end{document}